\documentclass{ws-ijgmmp}

\usepackage{mathtools}
\allowdisplaybreaks

\newcommand{\iN}{{\rm i}}
\newcommand{\eN}{{\rm e}}
\newcommand{\dimM}{\mathtt{D}}

\newcommand{\cofr}{\dform{\vartheta}}
\newcommand{\vfre}{\boldsymbol{e}}

\newcommand\dex{\mathrm{d}}
\newcommand\Dex{\mathbf{D}}
\newcommand\dint[1]{#1\rfloor}

\newcommand{\uU}[1]{\underline{U}^{#1}}
\newcommand{\uV}[1]{\underline{V}_{#1}}
\newcommand{\uW}[1]{\underline{W}_{#1}}
  
\newcommand{\kpar}{\mathtt{k}}

\newcommand{\dform}[1]{\boldsymbol{#1}}
\newcommand{\dfal}{\dform{\alpha}}
\newcommand{\dfvarDe}{\dform{\varDelta}}
\newcommand{\dfLa}{\dform{\varLambda}}
\newcommand{\dfPhi}{\dform{\Phi}}
\newcommand{\dfphi}{\dform{\phi}}
\newcommand{\dfPsi}{\dform{\Psi}}
\newcommand{\dfSi}{\dform{\Sigma}}
\newcommand{\dfga}{\dform{\gamma}}
\newcommand{\dfGa}{\dform{\Gamma}}
\newcommand{\dfb}{\dform{b}}
\newcommand{\dfh}{\dform{h}}
\newcommand{\dfk}{\dform{k}}
\newcommand{\dfm}{\dform{m}}
\newcommand{\dfq}{\dform{q}}
\newcommand{\dfA}{\dform{A}}
\newcommand{\dfE}{\dform{E}}
\newcommand{\dfF}{\dform{F}}
\newcommand{\dfH}{\dform{H}}
\newcommand{\dfK}{\dform{K}}
\newcommand{\dfL}{\dform{L}}
\newcommand{\dfM}{\dform{M}}
\newcommand{\dfP}{\dform{P}}
\newcommand{\dfQ}{\dform{Q}}
\newcommand{\dfR}{\dform{R}}
\newcommand{\dfT}{\dform{T}}
\newcommand{\dfW}{\dform{W}}
\newcommand{\dfX}{\dform{X}}
\newcommand{\dfY}{\dform{Y}}
\newcommand{\dfZ}{\dform{Z}}

\newcommand{\irrdfQ}[1]{{}^{\scriptscriptstyle(#1)\!}\dfQ}
\newcommand{\irrdfT}[1]{{}^{\scriptscriptstyle(#1)\!}\dfT}
\newcommand{\irrdfW}[1]{{}^{\scriptscriptstyle(#1)\!}\dfW}
\newcommand{\irrdfZ}[1]{{}^{\scriptscriptstyle(#1)\!}\dfZ}
\newcommand{\irrdfWLC}[1]{{}^{\scriptscriptstyle(#1)\!}\mathring{\dfW}}

\begin{document}

\markboth{Alejandro Jim\'enez-Cano}
{Review of gravitational wave solutions in quadratic metric-affine gravity}

%
\catchline{}{}{}{}{}
%

\title{REVIEW OF GRAVITATIONAL WAVE SOLUTIONS IN QUADRATIC METRIC-AFFINE GAUGE GRAVITY}

\author{ALEJANDRO JIM\'ENEZ-CANO}

\address{Laboratory of Theoretical Physics, Institute of Physics \\%
 University of Tartu, W. Ostwaldi 1, Tartu, 50411, Estonia\\
\email{alejandro.jimenez.cano@ut.ee}}

\maketitle

\begin{history}
\received{31.12.2021}
\revised{(Day Month Year)}
\end{history}

\begin{abstract}
In this review we consider the quadratic Metric-Affine Gauge gravity Lagrangian, which contains all the algebraic invariants up to quadratic order in torsion, nonmetricity and curvature. The goal will be to collect the known exact solutions for this theory that describe the propagation of a gravitational wave and some useful properties. We concentrate on solutions whose connections are different from the Levi-Civita one and provide also the extension of already known axial torsion solutions to more general Lagrangians. At the end, we will briefly discuss colliding waves and Riemannian solutions.
\end{abstract}


\section{Introduction}

The metric-affine framework is the result of considering, in addition to the metric, a general connection on the spacetime. This connection is treated as an independent field in the action and the equations of motion, and can potentially lead to new physics, contrasting with theories in which the connection is just an auxiliary field, such as Einstein-Palatini. If a metric is present, all  the information contained in the connection can be encoded in just two tensorial quantities, the torsion and the nonmetricity, defined in \eqref{eq:defTQR}-\eqref{eq:defTQR2}. These tensors are the building blocks to construct diffeomorphism-invariant Lagrangians in the metric-affine framework.

One of the most interesting features of this kind of geometry is that its objects are compatible with gauge formulations, in the sense that they can be derived as gauge connections for certain (non-linearly realized) symmetries; see for instance the explanations in \cite{PBO, JCAThesis}. One of the better-known examples is Poincaré Gauge gravity, a gauge theory from which a torsionful (but with zero nonmetricity) metric-affine structure arises \cite{HehlPG,ObuPG,BlagPG}. A similar procedure can be followed with the entire affine group and the result is a gauge theory in which both the torsion and the nonmetricity are non-trivial and (a priori) arbitrary. The latter is called Metric-Affine Gauge (MAG) gravity theory \cite{Hehl1995} (see also \cite{JCAThesis}), and its gauge fields are coupled to new matter charges  in addition to the energy momentum-tensor: the spin density, the shear and the dilation currents, all of them absent in General Relativity. These new currents are associated with what is commonly called in the literature the \emph{microstructure} of the matter \cite{matt1, matt2, matt3}. In addition to these, gauge theories of gravity with other groups can be built in the metric-affine framework (see \cite{gaugereader} and references therein). 

The Lagrangian of MAG is usually constructed as the most general combination up to quadratic terms in torsion, nonmetricity and curvature, providing dynamical terms for both the metric degrees of freedom and those of the connection. 
Nonetheless, this theory, as well as others in this same framework, is not free of problems. Due to the large number of new fields of different spin content and the intricate couplings of the action, pathologies such as ghosts, tachyons and strongly coupled backgrounds are expected (see {\it e.g.} \cite{Percacci,BelDelhom1,BelDelhom2,Marzo}). There are also many open questions and difficulties which, interestingly, are mostly related to the presence of the traceless part of the nonmetricity: construction of the generalization of the Dirac theory, understanding of the shear current (see {\it e.g.} \cite{shear}), presence of a spin-3 in the spectrum \cite{Percacci, Marzo}, the loss of the topological nature of Lovelock invariants in critical dimensions \cite{JJLov}, etc.

In addition to all this, as it is well known, exact solutions of a given theory encode information about its non-linear regime; moreover, stability criteria as well as the compatibility with observables allow to put constraints in the parameters of the action from which the solutions were derived. For these reasons, not only gravitational perturbations of a given theory around a certain background, but the search for exact gravitational wave solutions also plays an important role in gravitational physics, specially after the enormous boost of this field since the detection of the first gravitational wave signals \cite{gwdet1,gwdet2}. 

In this review we will revisit all\footnote{
    As far as the author is concerned, there are no other gravitational wave solutions of the quadratic MAG Lagrangian in the literature.} 
the known exact solutions of the quadratic MAG Lagrangian describing the propagation of a gravitational wave \cite{TuckerWang1995, Garcia2000, PuetzfeldDip, Puetzfeld2002, KingVassiliev2001, Vassiliev2002, Vassiliev2005, PasicVassiliev2005, PasicPhDThesis, PasicBarakovic2014, PasicBarakovic2015, Obukhov2006, ObukhovJCA2021}. Many of these solutions can be seen as generalizations or extensions of solutions in Poincaré gravity (see  \cite{PGwave1, PGwave2, PGwave3, PGwave4, PGwave5, PGwave6,PGwave7,PGwave8, PGwave9, PGwave10, PGwave11, PGwave12, PGwave13, PGwave14}) to the quadratic Lagrangian with non-trivial nonmetricity terms. It is worth highlighting that the present review focuses on those with non-trivial non-Riemannian structure, {\it i.e.}, those whose connection is not Levi-Civita. However, for completeness, at the end we will also briefly comment on colliding waves configurations found in \cite{Garcia1998, MaciasLammerzahl2000} and purely Riemannian solutions. 

Now we summarize the structure of this text. First, in Section \ref{sec:tools}, we present the metric-affine framework and all the tools needed to describe such a geometry (Subsection \ref{sec:MAG}) and then revise some generalities about the metric-affine dynamics and give the specific results for the quadratic MAG Lagrangian (Subsection \ref{sec:qMAG}). In Section  \ref{sec:waves}, we discuss the family of metrics we are interested in. Afterwards, the reader can find the wave solutions divided according to the different lines of research in which they were found. Regarding solutions with non-trivial matter, in Section \ref{sec:TW} and  Section \ref{sec:Puetz}, we will study those found in the presence of a spinor field \cite{TuckerWang1995} and the electrovacuum solutions in \cite{Garcia2000,Puetzfeld2002,PuetzfeldDip}, respectively. The following three sections concentrate on vacuum solutions. In Section \ref{sec:KV}, we cover the torsion waves in \cite{KingVassiliev2001,Vassiliev2002,Vassiliev2005} with Minkowski metric. Later, in Section \ref{sec:Pasic}, we continue with the results of \cite{PasicVassiliev2005, PasicPhDThesis, PasicBarakovic2014, PasicBarakovic2015}, who considered pp-wave metrics instead of the flat one and explored other torsion Ansatzes; in addition to this, we will show that the axial torsion solutions of \cite{PasicBarakovic2015} can be extended to more general curvature-square Lagrangians. Section \ref{sec:Obu} is devoted to the solutions in \cite{Obukhov2006, ObukhovJCA2021}, which include both torsion and nonmetricity for the general even-parity quadratic Lagrangian. We will finish the discussion of solutions with the miscellaneous Section \ref{sec:coll}, which contains other related cases: Riemannian solutions and the colliding waves in \cite{Garcia1998, MaciasLammerzahl2000}. We will conclude the review in Section \ref{sec:fin} by performing an overview of it, including  some additional comments. 

\subsection{Notation and conventions}

In this paper we collect the different results in unified notation and conventions. Therefore, the ones used here will not necessarily correspond to those used in the original publications. 

We choose natural units $c=\hbar=1$ and the convention $(+,-,-,-)$ for the Lorentzian signature. Indices are assumed to be raised/lowered with the spacetime metric. We will use Greek indices $\mu,\nu,\rho...$ for coordinate frames, whereas lowercase Latin indices will refer to a generic (usually non-holonomic) frame $\{\vfre_a\}$ and its dual coframe $\{\cofr^a\}$. The symbol $\dimM$ will represent the dimension of the manifold. In addition, we will abbreviate the exterior products of several coframes as $\cofr^{ab...c}\coloneqq \cofr^a\wedge\cofr^b\wedge...\wedge\cofr^c$ and the symmetrization and antisymmetrization of indices as $H_{(ab)}\coloneqq (H_{ab}+H_{ba})/2!$ and $H_{[ab]}\coloneqq (H_{ab}-H_{ba})/2!$, and similarly for more than two indices.

Regarding operations with differential forms, the symbol $\dint{\vfre_a}$ represents the interior product with respect to the vector field $\vfre_a$ (in particular, $\dint{\vfre_a}\cofr^b=\delta_a^b$). For the Hodge dual of a $p$-form $\dfal=\frac{1}{p!}\alpha_{a_1...a_p}\cofr^{a_1...a_p}$ we use the convention
\begin{equation}
  \star \dfal = \frac{1}{p!(\dimM-p)!} \alpha^{a_1...a_p} \mathcal{E}_{a_1...a_p a_{p+1}...a_\dimM}  \cofr^{a_{p+1}...a_\dimM}\,,
\end{equation}
where we have introduced the Levi-Civita tensor $\mathcal{E}_{a_1...a_\dimM}\coloneqq \sqrt{|\det(g_{ab})|}\epsilon_{a_1...a_\dimM}$ ($\epsilon_{1 2...\dimM}\equiv+1$), which is nothing but the tensorial components of the canonical volume form associated to the metric:
\begin{equation}
  \mathbf{vol}_{g} \coloneqq \sqrt{|\det(g_{\mu\nu})|}\dex x^1\wedge...\wedge x^\dimM = \frac{1}{\dimM!} \mathcal{E}_{a_1...a_\dimM} \cofr^{a_1...a_\dimM}.
\end{equation}
For the Levi-Civita connection we use the notation: $\mathring{\dfGa}_a{}^b$, $\mathring{\dfR}_a{}^b$...

Finally, just remark that overlined objects ($\overline{\mathcal{W}}$, $\overline{\mathcal{V}}$...) are odd-parity. An exception to this notation is Section \ref{sec:TW}, in which the line over a spinor will indicate the Dirac adjoint.

\section{Metric-Affine gravity}\label{sec:tools}

\subsection{The basic elements of Metric-Affine gravity}\label{sec:MAG}

In MAG, the information about the coordinate metric $g_{\mu\nu}$ is split into two objects, the (anholonomic) metric $g_{ab}$ and the coframe $\{\cofr^a= e_\mu{}^a \dex x^\mu\}$ (dual to a certain frame $\{\vfre_a =e^\mu{}_a \boldsymbol{\partial}_\mu \}$), which are indeed treated as independent fields. However, this independence is not real, since the anholonomic metric can always be reduced to the diagonal Minkowski form via a gauge fixing. Therefore, at the end of the day, the information contained in $g_{\mu\nu}$ is basically stored in the coframe. Having said this, the three basic objects of the geometry of MAG are the coframe $\{\cofr^a\}$, the metric $g_{ab}$ and the connection 1-form $\dfGa_a{}^b$. The exterior covariant derivative associated to $\dfGa_a{}^b$ of tensor-valued differential forms is given by
\begin{equation}
  \Dex \dfal^{a...}{}_{b...} \coloneqq \dex \dfal^{a...}{}_{b...} + \dfGa_c{}^a\wedge\dfal^{c...}{}_{b...} -\dfGa_b{}^c\wedge\dfal^{a...}{}_{c...} + \ldots\,
\end{equation}

It is possible to covariantize the exterior derivative of the three basic objects and the results are the torsion 2-form, the nonmetricity 1-form and the curvature 2-form. Respectively,
\begin{equation}
  \dfT^a\coloneqq \Dex \cofr^a,\qquad \dfQ_{ab}\coloneqq-\Dex g_{ab},\qquad \dfR_a{}^b \coloneqq \dex \dfGa_a{}^b + \dfGa_c{}^b\wedge\dfGa_a{}^c\,,\label{eq:defTQR}
\end{equation}
whose components are the well-known torsion, nonmetricity and curvature tensors:
\begin{equation}
  \dfT^a= \frac{1}{2} T_{\mu\nu}{}^a \dex x^\mu\wedge\dex x^\nu\,,\quad \dfQ_{ab}= Q_{\mu ab}\dex x^\mu\,,\quad\dfR_a{}^b= \frac{1}{2} R_{\mu\nu a}{}^b \dex x^\mu\wedge\dex x^\nu\,.\label{eq:defTQR2}
\end{equation}

Under the pseudo-orthogonal group, the torsion 2-form can be decomposed into three irreducible parts:
\begin{equation}
  \dfT^a=\irrdfT{1}{}^a+\irrdfT{2}{}^a+\irrdfT{3}{}^a\,.\label{eq:irredT}
\end{equation}
$\irrdfT{2}{}^a$ is the trace, $\irrdfT{3}{}^a$ contains the totally antisymmetric part of the torsion tensor and $\irrdfT{1}{}^a$ is the remaining tensorial part. On the other hand, the nonmetricity has four irreducible components:
\begin{equation}
  \dfQ_{ab}=\irrdfQ{1}{}_{ab}+\irrdfQ{2}{}_{ab}+\irrdfQ{3}{}_{ab}+\irrdfQ{4}{}_{ab}\,.
\end{equation}
Here $\irrdfQ{4}{}_{ab}$ is the trace in $ab$ of the nonmetricity (also known as the Weyl 1-form) and $\irrdfQ{3}{}_{ab}$ contains the other independent trace. The other two parts are traceless: $\irrdfQ{1}{}_{ab}$ is such that its components (with 3 indices) are totally symmetric, whereas $\irrdfQ{2}{}_{ab}$ represents just the remaining tensorial part.

For our purposes, it is enough to keep in mind that the trace parts $\irrdfT{2}{}^a$, $\irrdfQ{3}{}_{ab}$ and $\irrdfQ{4}{}_{ab}$ are completely determined by the 1-forms
\begin{align}
  \dfT &\coloneqq \dint{\vfre_a}\dfT^a &&= -T_{ac}{}^c\cofr^a\,,\\
  \dfLa&\coloneqq \cofr^a \dint{\vfre^b}(\dfQ_{ab}-g_{ab}\dfQ) &&=\left[Q^c{}_{ca}-\frac{1}{\dimM}Q_{ac}{}^c\right]\cofr^a \\
  \text{and}\qquad\dfQ &\coloneqq\frac{1}{\dimM} \dfQ_c{}^c&&=\frac{1}{\dimM}Q_{ac}{}^c\cofr^a\,,
\end{align}
respectively. The explicit expressions of the irreducible parts of $\dfT^a$ and $\dfQ_{ab}$, as well as the irreducible decomposition of the curvature can be found in  \ref{app:irred}. 

Consider a general MAG Lagrangian of the type
\begin{equation}
  \dfL = \dfL_\mathrm{grav} (g_{ab},\cofr^a,\dfQ_{ab},\dfT^a,\dfR_a{}^b) + \dfL_\mathrm{matt}(g_{ab},\cofr^a,\dfQ_{ab},\dfT^a,\dfR_a{}^b, \Psi,\Dex \Psi),
\end{equation}
where $\Psi$ represents a certain matter field. Thanks to the Noether identities under diffeomorphisms and general linear transformations, the equation of motion of $g_{ab}$ is redundant, and the connection and coframe equations are completely determined by the gravitational momenta
\begin{equation}
  \dfM^{ab} \coloneqq -2 \frac{\partial \dfL_\mathrm{grav}}{\partial \dfQ_{ab}}\,,\qquad \dfH_{a} \coloneqq - \frac{\partial \dfL_\mathrm{grav}}{\partial \dfT^a}\qquad\text{and} \qquad\dfH^a{}_b \coloneqq - \frac{\partial \dfL_\mathrm{grav}}{\partial \dfR_a{}^b}\,,
\end{equation}
and the matter currents (canonical energy-momentum and hypermomentum, respectively),
\begin{equation}
  \dfSi_a \coloneqq \frac{\delta S_\mathrm{matt}}{\delta \cofr^a}\qquad\text{and}\qquad \dfvarDe^a{}_b \coloneqq \frac{\delta S_\mathrm{matt}}{\delta \dfGa_a{}^b}\,.
\end{equation}
The explicit form of the equations are
\begin{align}
  0=\frac{\delta S}{\delta \cofr^a}&\equiv- \Dex \dfH_a  + \dfE_a + \dfSi_a \,,\label{eome}\\
  0=\frac{\delta S}{\delta \dfGa_a{}^b}&\equiv- \Dex \dfH^a{}_b  + \dfE^a{}_b + \dfvarDe^a{}_b\,.\label{eomw}
\end{align}
where
\begin{align}
\dfE_a &= \dint{\vfre_a} \dfL_\mathrm{grav} + (\dint{\vfre_a} \dfT^b)\wedge \dfH_b
+ (\dint{\vfre_a} \dfR_b{}^c)\wedge \dfH^b{}_c + \,{\frac 12}(\dint{\vfre_a} \dfQ_{bc})\dfM^{bc},\\
\dfE^a{}_b &= - \cofr^a \wedge \dfH_b - \dfM^a{}_b\,.
\end{align}

\subsection{The Lagrangian of quadratic Metric-Affine Gauge gravity}\label{sec:qMAG}

Since none of the solutions we are going to discuss appears in the context of a parity violating theory, we focus just on even-parity invariants to build the gravitational Lagrangian. Under that assumption, the most general one up to quadratic order in torsion, nonmetricity and curvature can be expressed
\begin{equation}
  \dfL_{\mathrm{MAG}}=\dfL_{\mathrm{w}}+\dfL_{\mathrm{s}}\,,\label{eq:LAG0}
\end{equation}
where the weak and strong gravitational parts are, respectively, defined as\footnote{Here we are using the same parameterization as \cite{ObukhovJCA2021}.}
\begin{align}
\dfL_{\mathrm{w}} & =-\frac{1}{2\kappa} \bigg[ 2\lambda \mathbf{vol}_{g} -a_{0}\dfR^{ab}\wedge\star\cofr_{ab}+\dfT^a\wedge\star\sum_{I=1}^{3}a_{I}\irrdfT{I}{}_{a}+\dfQ_{ab}\wedge\star\sum_{I=1}^{4}b_{I}\irrdfQ{I}{}^{ab}\nonumber \\
 & \qquad\qquad+2b_{5}(\irrdfQ{3}{}_{ac}\wedge\cofr^a) \wedge\star(\irrdfQ{4}{}^{bc}\wedge\cofr_{b})- 2\sum_{I=1}^{3}c_{I}\irrdfQ{I+1}{}_{ab}\wedge\cofr^a\wedge\star\dfT^b\bigg]\,,\label{eq:LagWeak} \\ 
\dfL_{\mathrm{s}} & = -\frac{1}{2\rho} \dfR_{ab}\wedge\star\Bigg[\sum_{I=1}^{6}w_{I} \irrdfW{I}{}^{ab}+v_{1}\cofr^a\wedge\big(\dint{\vfre_c}\irrdfW{5}{}^{cb}\big)\nonumber \\
 & \qquad\qquad\qquad +\sum_{I=1}^{5}z_{I}\irrdfZ{I}{}^{ab}+ v_{2}\cofr_c\wedge\big(\dint{\vfre^a} \irrdfZ{2}{}^{cb}\big)+\sum_{I=3}^{5}v_{I} \cofr^a\wedge\big(\dint{\vfre_c}\irrdfZ{I}{}^{cb}\big)\Bigg]\,.\label{eq:LagStrong}
\end{align}
This Lagrangian contains 28 dimensionless parameters ($a_I$, $b_I$, $c_I$, $w_I$, $z_I$, $v_I$) and one dimensionful parameter $\lambda$ (the cosmological constant term), together with the weak and the strong gravitational couplings, $\kappa$ and $\rho$, respectively.

Now let us express the equations of motion \eqref{eome}-\eqref{eomw} in a more convenient way. If we introduce the following decomposition of the momenta:
\begin{align}
\dfM^{ab} & \eqqcolon\frac{2}{\kappa}\dfm^{ab}\,,\\
\dfH_a & \eqqcolon\frac{1}{\kappa}\dfh_a\,,\\
\dfH^a{}_b & \eqqcolon-\frac{1}{2\kappa}a_0\star\cofr^a{}_b+\frac{1}{\rho}\dfh^a{}_b\,,\label{eq:decomHw}
\end{align}
it can be shown that in arbitrary dimension $\dimM$ these pieces are given by \cite{JCAThesis}
\begin{align}
\dfm^{ab} & =\star\Big\{\sum_{I=1}^{4}b_{I}\irrdfQ{I}{}^{ab}-b_{5} \Big[\cofr^{(a}(\dint{\vfre^{b)}}\dfQ) +\frac{1}{\dimM}g^{ab}(\dfLa-\dfQ)\Big]\nonumber \\
    & \qquad+c_{1}\dint{\vfre^{(a}}\dfT^{b)} -\frac{\dimM c_{1}-c_{2}-(\dimM-1)c_{3}}{\dimM(\dimM-1)} g^{ab}\dfT+\frac{c_{1}-c_{2}}{\dimM-1} \cofr^{(a}(\dint{\vfre^{b)}}\dfT)\Big\}\,,\label{eq: GMom g}\\
\dfh_a & =\star\Big\{ \sum_{I=1}^{3}a_{I}\irrdfT{I}{}_{a} +\cofr^b\wedge\sum_{I=2}^{4}c_{I-1} \irrdfQ{I}{}_{ab}\Big\}\,,\label{eq: GMom e}\\
\dfh_{ab} & = \star \Big\{ \sum_{I=1}^{6}w_{I}\irrdfW{I}{}_{ab}+\sum_{I=1}^{5}z_{I}\irrdfZ{I}{}_{ab}\nonumber \\
    & \qquad+\frac{1}{2}v_{1} \Big[\cofr_a\wedge(\dint{\vfre_c}\irrdfW{5}{}^c{}_{b})+\frac{1}{2}\cofr_{[a}\wedge\dint{\vfre_{b]}}\dfP\Big]\nonumber \\
    & \qquad+\frac{1}{2}v_{2} \Big[\cofr_c\wedge(\dint{\vfre_{(a}}\irrdfW{2}{}^c{}_{b)})+\cofr_c\wedge(\dint{\vfre_{[a}}\irrdfZ{2}{}^c{}_{b]})-2\irrdfZ{2}{}_{ab}\Big]\nonumber \\
    & \qquad+\frac{1}{2}v_{3} \Big[\cofr_a\wedge(\dint{\vfre_c}\irrdfZ{3}{}^c{}_{b})-\frac{1}{2}\cofr_{(a}\wedge(\dint{\vfre_{b)}}\dfP)+\frac{1}{\dimM}g_{ab}\dfP\Big]\nonumber \\
    & \qquad+\frac{1}{2}v_{4} \Big[\cofr_{[a}\wedge(\dint{\vfre_{|c|}}\irrdfZ{4}{}^c{}_{b]})+\cofr_{(a}\wedge(\dint{\vfre_{|c|}}\irrdfW{4}{}^c{}_{b)})+\dimM\irrdfZ{4}{}_{ab}\Big]\nonumber \\
    & \qquad+\frac{1}{2}v_{5} \Big[\cofr_a\wedge(\dint{\vfre_c}\irrdfZ{5}{}^c{}_{b})+\frac{1}{\dimM}g_{ab}\dfP\Big]\Big\}\,,\label{eq: GMom w}
\end{align}
where we are using the abbreviation
\begin{equation}
  \dfP \coloneqq \cofr^b \wedge (\dint{\vfre^a}\dfR_{ab}) \,. \label{eq:PPXYZ}
\end{equation}

If we consider the following splitting of the Lagrangian \eqref{eq:LAG0},
\begin{equation}
\dfL_{\mathrm{MAG}}=\frac{1}{2\kappa}\Big[a_{0}\dfR^{ab}\wedge\star\cofr_{ab}-2\lambda \mathbf{vol}_{g} \Big]+\frac{1}{\kappa}\dfL_{(2)}\,,
\end{equation}
then:\footnote{
   This result is valid in any dimension $\dimM$. Its extension including the odd parity sector in $\dimM=4$ can be found in \cite{JCAThesis}.}
\begin{theorem} \label{th:EoMqMAG} The variations with respect to the coframe and the connection of \textup{$S=\int\dfL_{\mathrm{MAG}}+S_{\mathrm{matt}}$} with $\dfL_{\mathrm{MAG}}$ given in \textup{\eqref{eq:LAG0}} and $S_{\mathrm{matt}}$ being a general matter action, can be written in the form
\begin{align}
\kappa\frac{\delta S}{\delta\cofr^a} & =\frac{a_{0}}{2}\dfR^{bc}\wedge\star\cofr{}_{bca} -\lambda\star\cofr_{a}+\dfq_{a}-\Dex\dfh_{a}+\kappa\dfSi_a\,, \label{eq:qMAGEqe0}\\
\kappa\frac{\delta S}{\delta\dfGa_a{}^b} & =\frac{a_{0}}{2} \left(\dfT^c\wedge\star\cofr{}^a{}_{bc} +\dfQ^{ac}\wedge\star\cofr{}_{cb}-\frac{1}{2}\dfQ_c{}^c\wedge\star\cofr{}^a{}_b\right)\nonumber \\
 &\quad -\cofr^a\wedge\dfh_{b} -2\dfm^a{}_{b} -\frac{\kappa}{\rho}\Dex\dfh^a{}_b+\kappa \dfvarDe^a{}_b \label{eq:qMAGEqw0}\,,
\end{align}
where
\begin{equation}
\dfq_a \coloneqq\dint{\vfre_{a}}\dfL_{(2)}\!+(\dint{\vfre_{a}}\dfT^{b})\wedge\dfh_{b}+(\dint{\vfre_{a}}\dfQ_{bc})\dfm^{bc}
 +\frac{\kappa}{\rho}(\dint{\vfre_{a}}\dfR_b{}^{c})\wedge\dfh^b{}_c\,.\label{eq:qdef}
\end{equation}
\end{theorem}

The gravitational sector of all the theories we are going to discuss are particular cases of the quadratic Lagrangian \eqref{eq:LAG0} with some extra matter Lagrangians. In particular, if we consider just the strong gravity sector (which will be the case in Section \ref{sec:KV} and Section \ref{sec:Pasic}), the only non-trivial contribution of the gravitational momenta comes from $\dfh_{ab}$ and the equations of motion (in vacuum) are:
\begin{align}
\frac{\delta S_{\mathrm{s}}}{\delta\cofr^a} & =\dint{\vfre_{a}}\dfL_{\mathrm{s}}\!
 +\frac{1}{\rho}(\dint{\vfre_{a}}\dfR_b{}^{c})\wedge\dfh^b{}_c\,, \label{eq:qMAGEqe0RR}\\
\rho\frac{\delta S_{\mathrm{s}}}{\delta\dfGa_a{}^b} & =- \Dex\dfh^a{}_b \label{eq:qMAGEqw0RR}\,.
\end{align}

\section{Useful results about Kundt spaces} \label{sec:waves}

The metric structures that we are going to explore in detail throughout this review belong to the so-called Kundt class \cite{Kundt} (see also \cite{Kundt2,Kundt3, Bicak}) and, to be precise, are subcases of \eqref{eq:genmet}. We start with the following definition:
\begin{definition}
 {\bf (Kundt space).} A \emph{Kundt space} is a Lorentzian manifold $(M, \boldsymbol{g})$ that admits a geodetic null congruence with vanishing optical scalars.
\end{definition}
The line element of a Kundt space in arbitrary dimensions $\dimM$ can always be written as (see {\it e.g.} \cite{Bicak})
\begin{equation}
  \dex s^2 =  H\dex u^2 + 2 \dex u \dex w +  2\mathcal{G}_A \dex u \dex x^A + g_{AB}\dex x^A\dex x^B\,, \label{eq:Kundt1}
\end{equation}
in appropriate coordinates $\{u, w, x^2,...,x^{\dimM-1} \}$, where $H=H(u, w, x^B)$, $\mathcal{G}_A=\mathcal{G}_A(u, w, x^B)$ and $g_{AB}=g_{AB}(u, x^B)$. Here $u$ and $w$ are null coordinates and the rest, which generate the so-called \emph{transversal space}, are spacelike. The uppercase Latin indices cover the $\dimM-2$ directions of the transversal space.

The velocity of the geodetic congruence that appears in the definition is called the \emph{wave vector} (field). It can be checked that it coincides with the coordinate field $\partial_w \eqqcolon k^\mu \partial_\mu$. Its dual, the \emph{wave form}, is given by
\begin{equation}
  \dfk \coloneqq k_\mu \dex x^\mu = \dex u\,.
\end{equation}

The Brinkmann spacetime \cite{Brink}, characterized by having a parallel (w.r.t. the Levi-Civita connection) null vector field, corresponds to the Kundt spaces \eqref{eq:Kundt1} after dropping the dependency on $w$. As a consequence, the wave vector becomes a Killing vector of the metric. It can be shown that, in such a case, the functions $\mathcal{G}_A$ can be absorbed in a coordinate transformation, so the metric takes the form
  \begin{equation}
    \dex s^2_\text{Brink} =  H(u,x^B)\dex u^2 + 2 \dex u \dex w +  g_{AB}(u,x^C)\dex x^A\dex x^B\,. \label{eq:Brink}
  \end{equation}

Henceforth in this manuscript we will work in a four dimensional manifold (we take $\dimM=4$). Under this restriction, we can enumerate some interesting properties of the Kundt spaces:
\begin{enumerate}
  \item The Kundt class includes in particular spacetimes of type-II, III and N in the Petrov classification, respecting the Pirani criterion for gravitational radiation \cite{Zakharov1973} (see also \cite{PuetzfeldDip, JCAThesis}). However, there are also Kundt spaces of type-D (included in the Pleba\'nski-Demia\'nski class) and conformally flat (type-O) \cite{Griff}.

  \item After an appropriate complex coordinate transformation $\{u,w,x^2,x^3\} \to \{u,w,\zeta,\zeta^*\}$ in the transversal space, the line element \eqref{eq:Kundt1} in four dimensions becomes (see \cite{Griff,Steph})
  \begin{equation}
    \dex s^2 =  \tilde{H}\dex u^2 +2 \dex u (\dex w + \mathcal{G} \dex \zeta + \mathcal{G}^* \dex \zeta^*) - \frac{2}{P^2} \dex \zeta \dex \zeta^* \,,\label{eq:Kundt2}
  \end{equation}
  where all the coordinates are null. Here we have introduced the real functions $\tilde{H}=H(u,w,x^B(\zeta,\zeta^*))$ and $P(u,\zeta,\zeta^*)$, and the complex function $\mathcal{G}(u,w,\zeta,\zeta^*)$.
\end{enumerate}

Now that we have introduced some general ideas about the Kundt class of metrics, we are going to focus on a particular sub-family of it. For our purposes, we are specially interested in the following metric
\begin{equation}
  \dex s^2 = 2 \dex u \left(\frac{Q}{P}\right)^2(S \, \dex u + \dex v) - \frac{2}{P^2} \dex \zeta \dex \zeta^*\,,\label{eq:genmet}
\end{equation}
where, again, we have taken a complex set of coordinates $(\zeta, \zeta^*)$ for the transversal space. The real functions $P$, $Q$ and $S$ are given by
\begin{align}
  P(\zeta,\zeta^*)&\coloneqq 1+ \frac{c}{6} \zeta \zeta^* ,\\
  Q(u, \zeta,\zeta^*)&\coloneqq \left(1- \frac{c}{6} \zeta \zeta^*\right)\alpha(u) + \zeta\beta^*(u)+\zeta^*\beta(u),\\
  S(u,v,\zeta,\zeta^*)&\coloneqq -v^2\left(\frac{c}{6}\alpha^2(u)+\beta(u)\beta^*(u)\right)+ v \partial_u \ln|Q|+ \frac{P}{2Q} \mathcal{H}(u,\zeta,\zeta^*),
\end{align}
for some arbitrary real functions $\mathcal{H}$ and $\alpha$, a certain complex function $\beta$ and a real constant $c$, which will be related to the cosmological constant in Section \ref{sec:Puetz}. The metric \eqref{eq:genmet} is indeed a Kundt metric that can be derived from \eqref{eq:Kundt2} by imposing the type-N condition and after the coordinate transformation $w=(Q/P)^2v$ with $P$ and $Q$ given by the previous expressions (see Section 18.3.2 of \cite{Griff}). 

The three irreducible parts of the Levi-Civita curvature, $\irrdfWLC{1}_{ab}$ (Weyl tensor), $\irrdfWLC{4}_{ab}$ (traceless Ricci tensor) and $\irrdfWLC{6}_{ab}$ (Ricci scalar), are non-trivial for non-vanishing $c$. In particular, the Ricci scalar is proportional to this constant,
\begin{equation}
  \irrdfWLC{6}_{ab}\propto\qquad\dint{\vfre_a}\dint{\vfre_b}\mathring{\dfR}{}^{ab}=-4c\,,
\end{equation} 
and is the only of the three that vanishes identically for $c=0$; the rest contain contributions proportional to the second derivatives of $\mathcal{H}$ with respect to the transversal coordinates. 

Now we turn our attention to gravitational wave criteria \cite{Zakharov1973} (see also \cite{PuetzfeldDip, JCAThesis}). Since, as we mention above, the metric \eqref{eq:genmet} is type-N, it verifies the Pirani criterion for gravitational radiation. Let us also consider the Lichnerowicz criterion, which consists on the following properties:
\begin{equation}
  \dfk\wedge\mathring{\dfR}_{ab} = 0\qquad\text{and} \qquad\dfk \wedge \star \mathring{\dfR}_{ab}=0\,.\label{eq:Lich}
\end{equation}
Respectively, in components,
\begin{equation}
  k_{[\mu}\mathring{R}_{\sigma\nu]ab} = 0\qquad\text{and} \qquad k^\mu\mathring{R}_{\mu\nu ab} = 0\,.
\end{equation}
A (tensor-valued) 2-form satisfies these two conditions if and only if it is the wedge product of $\dfk$ and a transversal (tensor-valued) 1-form \cite{JCA2020}.\footnote{
  By \emph{transversal} form here we mean that the \emph{internal} (hidden) indices are transversal. For instance, that the curvature is a transversal 2-form means that it has the form $\mathring{\dfR}_{ab}=\frac{1}{2}\mathring{R}_{ABab}\dex x^A\wedge \dex x^B$. Notice that this does not imply that the external indices $ab$ are transversal. Indeed, this is not the case as can be seen in the pp-wave case \eqref{eq:Rppwave}.}
As can be seen by direct computation, these conditions are not fulfilled for the Riemann curvature of \eqref{eq:genmet}, but the result is proportional to the constant $c$. Indeed, if we apply these radiation conditions to the remaining part of $\mathring{\dfR}_{ab}$ after removing the Ricci scalar,\footnote{
        In \cite{Garcia2000} the second condition was written as  $k^a(\mathring{\dfR}_{ab}-\irrdfWLC{6}_{ab})=0$, which is equivalent to the expression presented here thanks to the symmetry of the Riemann tensor under pair exchange: $\mathring{R}_{\mu\nu\rho\lambda}=\mathring{R}_{\rho\lambda\mu\nu}$.}
\begin{equation}
  \dfk\wedge(\mathring{\dfR}_{ab}-\irrdfWLC{6}_{ab}) = 0,\qquad \dfk \wedge \star (\mathring{\dfR}_{ab}-\irrdfWLC{6}_{ab})=0\,,\label{eq:Lich2}
\end{equation}
then they are fulfilled (see \ref{app:irred} for the explicit expression of $\irrdfW{6}_{ab}$).

Finally we will comment on a sub-case of \eqref{eq:genmet} that we will call \emph{pp-wave}. Such spaces are characterized by $c=0$, $\beta=0$ and $\alpha=1$ or, in other words, by $P=Q=1$ and $S=\frac{1}{2}\mathcal{H}(u,\zeta,\zeta)$,
\begin{equation}
  \dex s^2_\text{pp} = \mathcal{H}(u,\zeta,\zeta^*)\dex u^2 + 2 \dex u \dex v - 2\dex \zeta \dex \zeta^*\,.\label{eq:ppzeta}
\end{equation}
Notice that this metric is nothing but a Brinkmann space with flat transversal space ($g_{AB}=-\delta_{AB}$),
\begin{equation}
  \dex s^2_\text{pp} =  H(u,x,y)\dex u^2 + 2 \dex u \dex v - \dex x^2 - \dex y^2\,, \label{eq:ppw}
\end{equation}
where we have introduced the simplified notation $x^A=(x^2,x^3)\eqqcolon(x,y)$ for the transversal coordinates. By construction, these spaces are type-N, except for the degenerate case $H=1$ which is trivially type-O. The Levi-Civita curvature 2-form reads
\begin{equation}
  \mathring{\dfR}_{ab}= \partial_A\partial_B H\, \delta^A_{[a} k_{b]} \ \dex x^B\wedge \dfk\,,\label{eq:Rppwave}
\end{equation}
where the indices $a$ and $b$ are referred to an arbitrary coframe. As can be immediately deduced from the previous observations, the Ricci scalar irreducible part of this metric  ($\irrdfWLC{6}_{ab}$) is vanishing, whereas the traceless Ricci tensor and the Weyl tensor are non-trivial.

\section{Spinor-induced gravitational wave solutions}\label{sec:TW}

Here we follow the work \cite{TuckerWang1995} by R. W. Tucker and C. Wang. 

\subsection{Theory and  Ansatz}

Consider the Lagrangian
\begin{equation}
  \dfL^{\text{TW}}= \frac{a_0}{2\kappa} \dfR^{ab}\wedge\star\cofr_{ab} -\frac{z_5}{2\rho} \irrdfZ{5}{}_{ab}\wedge\star \irrdfZ{5}{}^{ab} \ +\ \dfL_{\text{D}}\,.\label{eq:TWLag}
\end{equation}
Here, we have introduced the (massless) Dirac Lagrangian in differential form notation 
\begin{equation}
  \dfL_{\text{D}}\coloneqq 
    - \frac{\iN}{2} \left( \overline \Psi\, \star\!\dfga \,\wedge\, \mathbb{D}\Psi +     \overline{\mathbb{D} \Psi}\, \wedge \, \star \dfga\, \Psi \right)\,,
\end{equation}
where $\overline \chi \coloneqq \chi^\dagger \gamma^0$ is the Dirac adjoint, $\dfga\coloneqq \gamma^a \cofr_a$ and a Clifford product should be understood between spinor-valued objects. Here, the spinor derivative is chosen to be
\begin{equation}
  \mathbb{D}\Psi \coloneqq \dex \Psi - \frac{1}{4} \dfGa_{ab}\, \gamma^{[a}\gamma^{b]}\Psi\,,
\end{equation}
and  $\{\gamma^a\}$ is a particular basis of the Clifford algebra satisfying the fundamental identity
\begin{equation}
  \gamma^{(a}\gamma^{b)}=g^{ab} \label{eq:gammaid}
\end{equation}
and the normalization
\begin{equation}
  \gamma^0\gamma^a\gamma^0= (\gamma^a)^\dagger\,. \label{eq:gammanor}
\end{equation}

The quadratic term in the segmental curvature $\irrdfZ{5}{}_{ab}$ is specially interesting because it serves as a kinetic term for the Weyl 1-form $\dfQ$ \cite{ObVlachy}. Indeed, according to \eqref{eq:Z5},
\begin{equation}
  \irrdfZ{5}{}_{ab}=\frac{1}{2}g_{ab}\dex \dfQ \qquad \Rightarrow\qquad \irrdfZ{5}{}_{ab}\wedge\star \irrdfZ{5}{}^{ab} = \dex \dfQ \wedge \star\dex \dfQ\,.
\end{equation}

Now that we know the Lagrangian, we can move on to describe the Ansatz. The authors considered a metric of the pp-wave type \eqref{eq:ppw} (via a orthonormal coframe).

For the connection, the only assumption is that the principal (Weyl) trace of the nonmetricity takes the form
\begin{equation}
  \dfQ = -\frac{1}{4} \big[ A(u) \dex x + B(u) \dex y\big]\,,\label{eq:TWQ}
\end{equation}
for certain real functions $A$ and $B$. As we will see, the rest of the pieces of the connection will be either set to zero or fixed by the equations of motion.

Finally, if we consider a basis in the spinor space such that the gamma matrices are of the form\footnote{We have multiplied the ones in the original paper by $\iN$ in order to respect our mostly minus convention.}
\begin{align}
\gamma^{0} =\begin{pmatrix} 0&0&\iN&0\\0&0&0&-\iN\\-\iN&0&0&0\\0& \iN&0&0\end{pmatrix},\qquad
\gamma^{1} =\begin{pmatrix} 0& 0 &\iN&0\\0&0&0&-\iN\\ \iN&0&0&0\\0&-\iN&0&0\end{pmatrix},&\nonumber\\
\gamma^{2} =\begin{pmatrix} 0&\iN&0&0\\\iN&0&0&0\\0&0&0&\iN\\0&0&\iN&0\end{pmatrix},\qquad
\gamma^{3}=\begin{pmatrix} 0&-1&0&0\\1&0&0&0\\0&0&0&-1\\0&0&1&0\end{pmatrix},&\label{eq:gammas}
\end{align}
the Dirac field is assumed to have the following components in such a basis:
\begin{equation}
  \Psi = ( 0, 0, \psi_3(u), \psi_4(u))\,, \label{eq:TWsp}
\end{equation}
for some complex functions $\psi_3(u)$ and $\psi_4(u)$. Notice that \eqref{eq:gammas} is consistent with \eqref{eq:gammaid} and \eqref{eq:gammanor}.

\subsection{Solutions}

The equations of motion of this theory are, in general,
\begin{align}
\kappa\frac{\delta S^\text{TW}}{\delta \cofr^a}=&&\frac{a_0}{2}\star\cofr_{abc}\wedge\dfR^{bc} + \kappa [\dfSi_a^{(\dfQ)}+\dfSi_a^{\text{D}}]& = 0\,,\label{eq:TWeqcof}\\
g_{ac}\frac{\delta S^\text{TW}}{\delta \dfGa_c{}^b}=&& \frac{a_0}{2\kappa}\left(\dfT^c\wedge \star \cofr_{abc}+\dfQ_a{}^c \wedge \star \cofr_{bc}-2\dfQ\wedge \star \cofr_{ab}\right)\nonumber\\
&&-\frac{z_5}{2\rho}g_{ab} \dex\star\dex\dfQ - \frac{\iN}{4} (\overline{\Psi} \gamma_{cab} \Psi)\star\cofr^c &=0 \,,\label{eq:TWeqcon}\\
  \frac{\delta S^\text{TW}}{\delta \overline\Psi}=&&\,-\iN \star \dfga \wedge \left(\mathbb{D}-\frac{1}{2}\dfM\right)\Psi\ &=0\,,\label{eq:Dirac} 
\end{align}
where
\begin{align}
  \dfSi_a^{(\dfQ)} & \coloneqq \frac{z_5}{2\rho}  \big[ (\dint{\vfre_a}\dex\dfQ) \wedge \star \dex\dfQ -(\dint{\vfre_a}\star\dex\dfQ) \wedge \dex\dfQ\,\big],\label{eq:SigQ}\\
  \dfSi_a^{\text{D}}&\coloneqq -\frac{\iN}{2} \left( \overline \Psi\, \star\!\dfga \ \mathbb{D}_a\Psi - 
    \overline{\mathbb{D}_a \Psi} \star\! \dfga\ \Psi \right)+\dint{\vfre_a}\dfL_{\text{D}},\qquad \mathbb{D}_a\coloneqq \dint{\vfre_a}\mathbb{D}\,,\\
  \dfM  & \coloneqq \dfT - \frac{1}{2}\dfLa + \frac{3}{2}\dfQ\,.
\end{align}

Let us now analyze each of the equations:
\begin{itemize}
\item The equation of the connection \eqref{eq:TWeqcon} is the same as in Einstein-Palatini gravity but with two extra terms: the one with $z_5$ and the one coming from the Dirac Lagrangian. The former is pure trace ($\propto g_{ab}$) and decouples from the rest, which is traceless. On the other hand, the Dirac contribution only enters the totally antisymmetric part of the equation in its three indices (once we remove the global $\star\cofr^c$). If we extract these two parts, {\it i.e.} the trace in $ab$ and the totally antisymmetric part of the equation in $cab$, we get the conditions:
\begin{align}
  \dex\star\dex\dfQ &=0 ,\\
  \irrdfT{3}_c &= \frac{\kappa}{4a_0} (\overline{\Psi}\iN\gamma_{abc} \Psi) \cofr^{ab}  \,.\label{eq:TWsolTor}
\end{align}
The first one is automatically fulfilled by the nonmetricity Ansatz \eqref{eq:TWQ} (thanks to the Hodge dual of the metric \eqref{eq:ppw}) for arbitrary $A$ and $B$. The second one fixes the totally antisymmetric part of the torsion $\irrdfT{3}^a$ in terms of the axial current.

Except for these two parts, the rest of the equation of the connection is the same as in Einstein-Palatini. The general solution is:
\begin{equation}
  \irrdfT{1}^a = \irrdfQ{1}_{ab} =\irrdfQ{2}_{ab}=\irrdfQ{3}_{ab}=0 ,\qquad   \irrdfT{2}^a = -\frac{1}{2} \cofr^a \wedge\dfQ  \,.
\end{equation}
Hence, the traces of the torsion and nonmetricity are constrained as follows: $\dfLa=0,\ \dfT = -\frac{3}{2}\dfQ$. Notice that this implies that the combination $\dfM$ that appears in the Dirac equation \eqref{eq:Dirac} vanishes. 

If we put all of this together we find that the general solution for the connection is
\begin{equation}
  \dfGa_{ab} = \mathring{\dfGa}_{ab} + \frac{1}{2}g_{ab}\dfQ + \frac{\kappa}{4a_0} (\overline{\Psi} \iN\gamma_{abc} \Psi) \cofr^c\,.\label{eq:TWconSOL}
\end{equation}

\item If we use the solution of the equation of the connection \eqref{eq:TWconSOL} as well as the Levi-Civita connection of the metric, and plug the Ansatz \eqref{eq:TWsp} for the spinor, the Dirac equation \eqref{eq:Dirac} is identically satisfied for generic $\psi_3$ and $\psi_4$.

\item It can be shown that if the connection and the matter equations are satisfied, only the symmetric part of the equation of the coframe \eqref{eq:TWeqcof} is non-trivial. This equation, once we express it in the coordinate basis $\{u,v,x,y\}$,  only has one nontrivial component, which leads to the following differential equation for $H$
\begin{equation}
  \frac{a_0}{2}(\partial_x^2+\partial_y^2)H=\frac{\kappa}{16\rho}z_5\big[(A')^2 + (B')^2\big] + \kappa {\rm Im} (\psi_3^*\partial_u\psi_3 + \psi_4^*\partial_u\psi_4 )\,.\label{eq:EQH}
\end{equation}
\end{itemize}
Observe that the variables $A$, $B$, $\psi_3$ and $\psi_4$ remain unfixed. Now we can collect the results of \cite{TuckerWang1995} in the following theorem:

\begin{theorem}
A metric-affine geometry described by a pp-wave metric \eqref{eq:ppw} and a connection satisfying \eqref{eq:TWQ}, together with the propagating spinor field given by \eqref{eq:TWsp} (in the basis in which \eqref{eq:gammas} is true) is a solution of the equations of motion of the theory \eqref{eq:TWLag} if and only if:
\begin{enumerate}
\item only the principal (Weyl) trace of the nonmetricity is non-vanishing ($\dfQ_{ab}=\irrdfQ{4}_{ab}$);
\item the torsion is purely axial ($\dfT^a=\irrdfT{3}^a$) and given by \eqref{eq:TWsolTor} in terms of the axial current of the Dirac spinor;
\item and, for given $A$, $B$, $\psi_3$ and $\psi_4$, the differential equation \eqref{eq:EQH} is fulfilled by the function $H$.
\end{enumerate}
\end{theorem}

\section{Electrovacuum solutions}\label{sec:Puetz}

In this section we analyze the solutions by A. Garc\'ia, A. Mac\'ias, D. Puetzfeld and J. Socorro \cite{Garcia2000}, which are also described in \cite{PuetzfeldDip, Puetzfeld2002}.

\subsection{Theory and  Ansatz}

They considered the gravitational Lagrangian obtained from \eqref{eq:LAG0} after switching off the entire quadratic sector in curvature except the term with $z_5$, and coupled it to the standard Maxwell term,
\begin{equation}
  \dfL^{\text{GMPS}}= \dfL_{\mathrm{w}} -\frac{z_5}{2\rho} \irrdfZ{5}{}_{ab}\wedge\star \irrdfZ{5}{}^{ab} \ -\ \frac{1}{2} \dfF\wedge\star\dfF\,.\label{eq:Gar}
\end{equation}
No other matter fields were added, since the aim was to explore electrovacuum solutions.

The Ansatz for the geometry can be constructed as follows. The line element is taken to be of the type-N Kundt form \eqref{eq:genmet} with $c=\lambda$, the cosmological constant. For the connection the authors considered the so-called {\it triplet Ansatz}, which has been shown to be a very useful technique to find solutions in MAG theories (see {\it e.g.} \cite{Puetzfeld2002, ObVlachy, GarciaPleban1998, Vlachy1996, ObuCharges}). This triplet Ansatz basically means that the nonmetricity and the torsion are assumed to be pure-trace,
\begin{equation}
  \dfQ_{ab}=\irrdfQ{3}_{ab}+\irrdfQ{4}_{ab}\,,\qquad   \dfT^a=\irrdfT{2}^a\,,
\end{equation}
and such that the three 1-forms that determine the non-Riemannian part of the connection are proportional to each other, {\it i.e.}
\begin{equation}
  \dfQ= \kpar_0 \dfphi ,\qquad \dfLa = \kpar_1 \dfphi, \qquad \dfT= \kpar_2 \dfphi\,, \label{eq:triplet}
\end{equation}
where $\dfphi$ is a certain 1-form (to be determined) and $\kpar_0$, $\kpar_1$ and $\kpar_2$ are real parameters given by
\begin{align}
  \kpar_0 &\coloneqq 4(a_2-2a_0)\left(\frac{a_0}{8}+b_3\right)-3(a_0-c_2)^2\,,\\
  \kpar_1 &\coloneqq \frac{9}{2}(a_2-2a_0)(2b_5-a_0)-9(a_0-c_2)(a_0-c_3)\,,\\
  \kpar_2 &\coloneqq 12\left(\frac{a_0}{8}+b_3\right)(a_0-c_3)-\frac{9}{2} (2b_5-a_0)(a_0-c_2) \,.\label{eq:tripletF}
\end{align}

~

In addition to the previous Ansatz, the authors also imposed the set of conditions:
\begin{equation}
  a_0=1,\qquad z_5=\frac{\rho}{\kpar_0^2}\,,\label{eq:Garciacond}
\end{equation}
\begin{equation}
  (m^2 \coloneqq )\quad\frac{-1}{z_5 \kappa} \left(-4b_4+3\frac{a_0}{2}+\frac{\kpar_1}{2\kpar_0}(2b_5-a_0)+\frac{\kpar_2}{\kpar_0}(-c_3+a_0) \right)=0\,.\label{eq:m2}
\end{equation}
The following Ansatz for the electromagnetic 2-form is assumed:
\begin{equation}
  \dfF = \big[\, f(\zeta,u) \dex \zeta + f^*(\zeta^*,u) \dex \zeta^* \, \big] \wedge \dex u\,,\label{eq:GarciaF}
\end{equation}
which fulfills the radiation conditions $\dfk\wedge\dfF=0$ and $\dfk\wedge\star\dfF=0$ (with $\dfk=\dex u$) and is compatible ($\dfF=\dex\dfA$) with the electromagnetic potential
\begin{equation}
  \dfA = \left( \int^\zeta f(\sigma,u)\dex \sigma + \int^{\zeta^*} f^*(\sigma^*,u)\dex \sigma^*\right)\wedge \dex u\,.\label{eq:GarciaA}
\end{equation}

Motivated by the equations of motion (see below) the authors also assumed a structure of the type \eqref{eq:GarciaA} for $\dfphi$, but depending on a different complex function $g(\zeta, u)$:
\begin{equation}
  \dfphi = \left( \int^\zeta g(\sigma,u)\dex \sigma + \int^{\zeta^*} g^*(\sigma^*,u)\dex \sigma^*\right)\wedge \dex u\,.\label{eq:Garciafi}
\end{equation}

\subsection{Solutions}
For the considered Lagrangian, the dynamical equations under the triplet Ansatz lead us to the Einstein equation (with cosmological constant) coupled to a Maxwell and a Proca energy-momentum sources:\footnote{
    The fact that the equations of motion of the gravitational sector of \eqref{eq:Gar} lead to the Einstein-Proca dynamics under the triplet Ansatz is known as Obukhov's equivalence theorem. See {\it e.g.} \cite{Hehl1999exact} and the original publication \cite{ObVlachy}.}
\begin{align}
\kappa\frac{\delta S^\text{GMPS}}{\delta \cofr^a}|_\text{Triplet}=&&\frac{a_0}{2}\star\cofr_{abc}\wedge\mathring{\dfR}^{bc} - \lambda \star\cofr_a + \kappa [\dfSi_a^{(\dfphi)}+\dfSi_a^{\text{Max}}]& = 0\,,\label{eq:Geqcof}\\
g_{ac}\frac{\delta S^\text{GMPS}}{\delta \dfGa_c{}^b}|_\text{Triplet}=&& -\frac{z_5 \kpar_0}{2\rho}g_{ab} (\dex\star\dex\dfphi+m^2\star\dfphi) &=0 \,,\label{eq:Geqcon}\\
  \frac{\delta S^\text{GMPS}}{\delta \dfA}=&&\,-\dex\star\dfF &=0\,,\label{eq:GMax} 
\end{align}
where the mass parameter $m^2$ (see \cite{ObVlachy}) was defined in \eqref{eq:m2} and
\begin{align}
  \dfSi_a^{(\dfphi)}   &\coloneqq \frac{z_5\kpar_0^2}{2\rho} \Big\{ (\dint{\vfre_a}\dex\dfphi) \wedge \star \dex\dfphi -(\dint{\vfre_a}\star\dex\dfphi) \wedge \dex\dfphi \\
  &\qquad\qquad+ m^2 [(\dint{\vfre_a}\dfphi) \wedge \star \dfphi +(\dint{\vfre_a}\star\dfphi) \wedge \dfphi]\Big\}\,,\\
\dfSi_a^{\text{Max}} &\coloneqq \frac{1}{2} \left[(\dint{\vfre_a}\dex\dfA) \wedge \star \dex\dfA -(\dint{\vfre_a}\star\dex\dfA) \wedge \dex\dfA \right]\,.
\end{align}

\begin{itemize}
\item The Maxwell equation \eqref{eq:GMax} is automatically verified by \eqref{eq:GarciaF}.

\item After imposing the vanishing of the mass, \eqref{eq:m2}, the equation of the connection \eqref{eq:Geqcon} reduces to the Maxwell equation. Therefore it seems reasonable to assume \eqref{eq:Garciafi}, which is a solution of it.

\item After using all the previous results and \eqref{eq:Garciacond}, one can check that the equation of the coframe \eqref{eq:Geqcof} gives just the following condition:
\begin{equation}
  \partial_{\zeta}\partial_{\zeta^*}\mathcal{H} + \frac{\lambda}{3P^2}\mathcal{H} = \frac{2\kappa P}{Q} (|f|^2 +|g|^2)\,.\label{eq:EQHGar}
\end{equation}
This is a linear differential equation for $\mathcal{H}$, whose solutions can be computed. We omit here the explicit results that can be found in the original publication \cite{Garcia2000}, in which, additionally, the solutions for $\alpha(u) =1$ and $\beta(u) = 0$ are described for the particular choice $f=f_0 \zeta ^n$ and $g = g_0 \zeta ^l$ with $n,l\in\mathbb{Z}$.
\end{itemize}
We can finally conclude:
\begin{theorem}
A metric-affine geometry described by a Kundt space of the type \eqref{eq:genmet} and a triplet connection \eqref{eq:triplet}-\eqref{eq:tripletF} such that the 1-form $\dfphi$ is given by \eqref{eq:Garciafi}, together with the Maxwell field given by \eqref{eq:GarciaA} is a solution of the equations of motion of \eqref{eq:Gar} if and only if:
\begin{enumerate}
\item the parameters of the theory satisfy the constraints \eqref{eq:Garciacond} and \eqref{eq:m2};
\item and, for given $f$ and $g$, the differential equation \eqref{eq:EQHGar} is fulfilled by $\mathcal{H}$.
\end{enumerate}
\end{theorem}

\section{Torsion waves and pseudo-instantons}\label{sec:KV}

Now we concentrate in the results by A. D. King and D. Vassiliev in \cite{KingVassiliev2001, Vassiliev2002, Vassiliev2005}.

\subsection{Theory and Ansatz}
In these works there is a progressive generalization of the Lagrangian. In the latter work \cite{Vassiliev2005}, the author considered the general Lagrangian quadratic in curvature, {\it i.e.},  the strong sector \eqref{eq:LagStrong}. In \cite{Vassiliev2002}, only the pure squares of the irreducible parts were considered  ($v_I=0$),
\begin{equation}
  \dfL^\text{V}= -\frac{1}{2\rho} \dfR_{ab}\wedge\star\Bigg[\sum_{I=1}^{6}w_{I} \irrdfW{I}{}^{ab}+ \sum_{I=1}^{5}z_{I}\irrdfZ{I}{}^{ab}\Bigg]\,.\label{eq:LagV}
\end{equation}
Finally, in the first work of this line of research, \cite{KingVassiliev2001},  the reduced case $w_I=-1, z_I=1$ was considered. In this case, the Lagrangian can be re-expressed as
\begin{equation}
  \dfL^\text{KV}= -\frac{1}{2\rho} \dfR_a{}^b\wedge\star\dfR_b{}^a\,, \label{eq:LagKV}
\end{equation}
which is similar to the standard Yang-Mills Lagrangian.

Regarding the Ansatzes in \cite{KingVassiliev2001, Vassiliev2002, Vassiliev2005} in which we are interested ({\it i.e.}, the torsion waves), the metric is assumed to be Minkowski in diagonal form,
\begin{equation}
  \dex s^2_\text{Mink}= \dex t^2 - \dex x^2- \dex y^2- \dex z^2.
\end{equation}
We can then take the coframe to be just the Cartesian basis $\{\dex x^\mu\}$ (and, therefore, $g_{\mu\nu}\equiv g_{ab}$). In addition to this, since the connection is chosen to be metric-compatible ($\dfQ_{ab}=0$), we conclude that the only non-trivial information about the geometry is contained in the torsion.

To be precise, the contorsion 1-form $\dfK_a{}^b$, defined implicitly by $\dfT^a \eqqcolon \dfK_b{}^a\wedge\cofr^b$,
is assumed to be of the form 
\begin{equation}
  \dfK_{ab} = \mathrm{Re}\left(\dfb L_{ab}\right)\,,\label{eq:KVcont}
\end{equation} 
where $L_{ab}=L_{[ab]}$ is a constant complex (tensor-valued) 0-form verifying
\begin{equation}
  \star(L_{ab}\cofr^{ab})=\tilde\alpha \iN\ L_{ab}\cofr^{ab},\qquad \tilde\alpha\in\{-1,1\}, \label{eq:Leigen}
\end{equation} 
and $\dfb$ is a complex 1-form fulfilling the {\it polarized} Maxwell equation
\begin{equation}
  \star\dex\dfb = \alpha \iN\ \dex \dfb,\qquad \alpha\in\{-1,1\}.\label{eq:KVpolMax}
\end{equation} 
Note that a solution of the latter automatically verifies the standard Maxwell equation $\dex\star\dex\dfb =0$. In particular, we focus on plane-wave solutions of \eqref{eq:KVpolMax}. Up to a proper Lorentz transformation \cite{KingVassiliev2001}, they are given by
\begin{equation}
  \dfb = C \eN^{\pm \iN (t+z)} (\dex x - \iN \alpha \dex y)\,,\qquad C\in\mathbb{R}.\label{eq:KVwave}
\end{equation}

\subsection{Solutions}

In \cite{KingVassiliev2001} the authors showed that the Ansatz described above is indeed a solution of the theory \eqref{eq:LagKV}. In order to restrict the set of vacuum solutions, the authors imposed
\begin{equation}
\dint{\vfre_a}\dfR_b{}^a=0\,,
\end{equation}
which is an Einstein-like equation but using the Ricci form $\dint{\vfre_a}\dfR_b{}^a$ of the connection $\dfGa_a{}^b$ (not only the Levi-Civita part), and found that 
\begin{equation}
L_{ab} = (\dint{\vfre_b}\dint{\vfre_a}\dex \dfb)|_{t=x=y=z=0}\,.\label{eq:KVL}
\end{equation}
The resulting torsion is completely determined by the propagating form \eqref{eq:KVwave} and is purely tensorial ($\dfT^a=\irrdfT{1}^a$), {\it i.e.}, the trace and the totally antisymmetric parts of the torsion tensor vanish. By hypothesis from the beginning, $L_{ab}$ is a constant object; however, the condition \eqref{eq:KVL} can be rewritten as $L_{ab} = (\dint{\vfre_b}\dint{\vfre_a}\dex \dfb)$ after an appropriate rescaling of the wave vector $k^\mu=(1,0,0,1)$ (see \cite{KingVassiliev2001}), allowing to express the contorsion as
\begin{equation}
  (e_\mu{}^ce_\nu{}^a e_\rho{}^b\dint{\vfre_c}\dfK_{ab}\eqqcolon)\qquad K_{\mu\nu\rho} = 2\mathrm{Re}\left(b_\mu \partial_{[\nu} b_{\rho]} \right). \label{eq:Ldb}
\end{equation}

Interestingly, the structure of the solution can be re-interpreted by using spinor language. The authors showed that there is a one-to-one correspondence between equivalence classes of solutions $\dfb$ (under $\dfb\sim\dfb'$ iff $\dex\dfb=\dex\dfb'$) and those of the massless Dirac equation together with another condition over the spinor (see Lemma 8 in \cite{KingVassiliev2001} for more details). It is quite remarkable that the sign $\alpha$ introduced in \eqref{eq:KVpolMax} ends up distinguishing (via this mapping) between left-handed and right-handed spinor solutions.

~

The second work \cite{Vassiliev2002} is focused on the Lagrangian \eqref{eq:LagV} with arbitrary weights $w_I$ and $z_I$. The author defined a {\it pseudo-instanton} as any configuration for the connection with zero nonmetricity and purely irreducible curvature. Interestingly, the main theorem of \cite{Vassiliev2002} says that any pseudo-instanton is a solution of the dynamical equations of \eqref{eq:LagV}. In this work, it is also shown that the wave Ansatz defined by \eqref{eq:KVwave} and \eqref{eq:Ldb} (with $\dfQ_{ab}=0$ and $\cofr^a= \delta^a_\mu \dex x^\mu$) constitutes a solution of the general Lagrangian \eqref{eq:LagV}. This is indeed, a direct consequence of the theorem mentioned above, since the corresponding curvature coincides with its Weyl irreducible part ($\dfR_a{}^b = \irrdfW{1}_a{}^b$) and, therefore, the geometry is a Weyl pseudo-instanton.

~

Finally, in \cite{Vassiliev2005}, it is also proven that a pseudo-instanton fulfills the dynamical equations of the general strong sector \eqref{eq:LagStrong}. However it is important to notice that the notion of pseudo-instanton used in this paper is \emph{more restrictive} than the one of \cite{Vassiliev2002}. To the previous definition, we have to add that the curvature must be not only irreducible but also {\it simple}, which means that the irreducible part cannot be isomorphic to any other. This constrains the possibilities, since among the six irreducible parts of the curvature of a metric-compatible connection, only three are simple. It can be seen that these actually coincide with the ones that are non-vanishing in the Levi-Civita case: the Weyl, the traceless symmetric-Ricci and the Ricci scalar parts. Since the curvature of the configuration considered in the previous paper \cite{Vassiliev2002} is purely Weyl, it also respects this second definition of pseudo-instanton and, therefore, it must be a solution for the complete strong sector \cite{Vassiliev2005}.

~

We have seen that the result of \cite{Vassiliev2005} generalizes the solution in \cite{Vassiliev2002} to the general second-order Lagrangian in curvature. Now we collect the results of this section in the following theorem:
\begin{theorem}
Consider a metric-affine geometry whose metric is Minkowski and whose connection is metric-compatible with contorsion of the form \eqref{eq:KVcont}, such that \eqref{eq:Leigen} and \eqref{eq:KVwave} are true. Then: 
\begin{enumerate}
\item For this geometry, the equations of motion of \eqref{eq:LagKV} are satisfied.
\item Under the additional restriction \eqref{eq:Ldb}, the geometry is a Weyl pseudo-instanton (according to both definitions, the one in \textup{\cite{Vassiliev2002}} and the one in \textup{\cite{Vassiliev2005}}) and the equations of motion of the general strong sector \eqref{eq:LagStrong} are satisfied.
\end{enumerate}
\end{theorem}

\section{Other torsion wave solutions}\label{sec:Pasic}

In this section we collect later results by V. Pasic, D. Vassiliev and E. Barakovic along the line of research described in the previous section. In these works, generalized metrics were consider \cite{PasicVassiliev2005, PasicPhDThesis, PasicBarakovic2014} as well as alternative Ansatzes for the torsion \cite{PasicBarakovic2015} (see also \cite{Pasic2017}).

\subsection{Solutions with pp-wave metric and purely tensor torsion}

In \cite{PasicVassiliev2005} (see also \cite{PasicPhDThesis}), the authors continue working with the complete strong sector of the quadratic MAG Lagrangian but generalize the metric from Minkowski to a pp-wave  \eqref{eq:ppw}. They again considered a metric-compatible connection with contorsion \eqref{eq:Ldb} and with $\dfb$ being a plane-wave solution of \eqref{eq:KVpolMax}, where now the Hodge dual is the one of the pp-wave metric.\footnote{
    The authors of \cite{PasicVassiliev2005} call a complex vector field $b^\mu$ (or the associated 1-form) {\it plane wave} if  $V^\nu\mathring{\nabla}_\nu b^\mu=0$ holds for all vector fields $V^\nu$ orthogonal to the wave vector $k^\mu$.} 
Such solutions were presented in \cite{PasicVassiliev2005} and have the form:
\begin{equation}
  \dfb = G_1(\varphi) \dex u - G_2(\varphi) (\dex x - \alpha \iN \dex y)\,,\label{eq:bBrink}
\end{equation}
for some functions $G_1$ and $G_2$ and where the phase of the wave is $\varphi=u+$constant. There are some interesting features of this geometry:
\begin{itemize}
\item The torsion continues being purely tensorial ($\dfT^a=\irrdfT{1}^a$).
\item The curvature has the same irreducible components as the Levi-Civita one and, indeed, the non-Riemannian contributions only enter the Weyl part $\irrdfW{1}_{ab}$. In other words, the traceless-Ricci scalar part $\irrdfW{4}_{ab}$ and the Ricci scalar part $\irrdfW{6}_{ab}$ are those of Levi-Civita. In particular, the latter is vanishing.
\item There is no contribution of the non-Riemannian part to the covariant derivative of the Ricci tensor, {\it i.e.}, $\nabla \text{Ric}=\mathring{\nabla} \text{Ric}$. Moreover, thanks to the previous observation, there is no distinction between the Ricci tensor of the connection $\dfGa_a{}^b$ and the Levi-Civita one, so
\begin{equation}
  \nabla {\rm Ric}=\mathring{\nabla} {\rm Ric}= \nabla \mathring{{\rm Ric}}=\mathring{\nabla} \mathring{{\rm Ric}}\,.\label{eq:derricci}
\end{equation}
The Ricci tensor is parallel for the pp-wave geometry essentially if $H$ is harmonic in the transversal space, {\it i.e.},
\begin{equation}
  (\partial_x^2+\partial_y^2)H=0.\label{eq:derriccihar}
\end{equation}

\end{itemize}
In \cite{PasicVassiliev2005} (see also \cite{PasicPhDThesis}) the following result is presented:
\begin{theorem}
Consider a metric-affine geometry described by a pp-wave metric \eqref{eq:ppw} and a metric-compatible connection whose contorsion is of the form \eqref{eq:Ldb} with $\dfb$ a solution of \eqref{eq:KVpolMax} ({\it i.e.}, it is given by \eqref{eq:bBrink}).  If the Ricci tensor is parallel \eqref{eq:derriccihar}, the equations of motion of the general strong sector \eqref{eq:LagStrong} are satisfied.
\end{theorem}

Later in \cite{PasicBarakovic2014} some errors in the dynamical equations that do not affect the solutions were noticed and corrected. In this work, the authors also discuss extensively about the interpretation of the solutions and their similarity with pp-wave solutions of Einstein gravity coupled to a massless spinor field.

\subsection{pp-wave metric with purely axial torsion: new solutions}

In \cite{Pasic2017}, the explicit form of the equations of motion of \eqref{eq:LagStrong} was given for metric-compatible spacetimes satisfying the conditions
\begin{equation}
  R_{[abcd]}=0,\qquad R_{ab}{}^{ab}=0\,,\qquad R_{abcd}=R_{cdab}.\label{eq:condR}
\end{equation}
In terms of irreducible components, due to the metric-compatibility, all the $\irrdfZ{I}_{ab}$ vanish. The first two conditions in \eqref{eq:condR} are equivalent to, respectively $\irrdfW{3}_{ab}=0$ and $\irrdfW{6}_{ab}=0$. Moreover, the last condition in \eqref{eq:condR} implies $\irrdfW{2}_{ab}=-\irrdfW{5}_{ab}$, so the curvature can be expanded just as
\begin{equation}
  \dfR_{ab} = \irrdfW{1}_{ab}+\irrdfW{4}_{ab}\,.
\end{equation}
Notice that this does not imply that $\irrdfW{2}_{ab}=\irrdfW{5}_{ab}=0$. In \cite{Pasic2017}, the authors also presented the equations for the particular case in which the torsion is purely axial.

In the work \cite{PasicBarakovic2015}, solutions with axial torsion were considered for the Yang-Mills-like restriction \eqref{eq:LagKV} of the strong sector. The metric is again assumed to be a pp-wave \eqref{eq:ppw}, whereas for the connection we take $\dfQ_{ab}=0$ and
\begin{equation}
  \dfT^a = \irrdfT{3}^a = \frac{1}{3} \dint{\vfre^a}\star\overline{\dfT}.\label{eq:PasicT}
\end{equation}
The latter can be equivalently written
\begin{equation}
  T_{abc}\cofr^{abc} = \dfT_a\wedge\cofr^a= \star \overline{\dfT}\,,
\end{equation}
as it was presented in \cite{PasicBarakovic2015}. Here, the axial 1-form $\overline{\dfT}$ is of the form
\begin{equation}
  \overline{\dfT}=\mathcal{A}(\varphi)\dex u\,, \label{eq:PasicTax}
\end{equation}
with $\mathcal{A}$ an arbitrary function. Comparing with \eqref{eq:bBrink}, we realize that this object satisfies the Maxwell polarized equation $\star \dex \overline{\dfT} = \pm \iN \dex \overline{\dfT}$. The full torsion 2-form reads
\begin{equation}
  \dfT_a=\frac{1}{6} \mathcal{A}(\varphi) k^d \mathcal{E}_{dbca} \cofr^{bc}\,.
\end{equation}

The irreducible components $\irrdfW{I}_{ab}$ with $I=1,2,3,4$ are non-trivial for this geometry, and only the cases $I=2,3,4$ receive non-Riemannian corrections. In particular, the Ricci tensor is of the form
\begin{equation}
  {\rm Ric}= \mathring{{\rm Ric}} + \frac{1}{18} \mathcal{A}(\varphi)^2 \dex u\otimes\dex u\,,
\end{equation}
where we are using the convention ${\rm Ric}\coloneqq R_{acb}{}^c\cofr^a\otimes\cofr^b$. In this case, we again have the property that the full covariant derivative and the Levi-Civita one coincide when acting on the Ricci tensors (but separately since these do not coincide for this geometry)
\begin{equation}
  \nabla {\rm Ric}=\mathring{\nabla} {\rm Ric},\qquad \nabla \mathring{{\rm Ric}}=\mathring{\nabla} \mathring{{\rm Ric}}\,.
\end{equation}
With this in mind, it can be shown that the Ricci tensor of the torsionful connection, ${\rm Ric}$, is parallel if and only if
\begin{equation}
  (\partial_x^2+\partial_y^2)H=\frac{1}{9} \mathcal{A}(\varphi)^2+ C,\qquad C\in\mathbb{R}\,.\label{eq:Ricparalaxial}
\end{equation}
Interestingly, this condition lead us to a new family of solutions \cite{PasicBarakovic2015}:
\begin{theorem}\label{thm:Ax}
Consider a metric-affine geometry described by a pp-wave metric \eqref{eq:ppw} and a metric-compatible connection with purely axial torsion given by \eqref{eq:PasicT} and \eqref{eq:PasicTax}. Assume that the Ricci tensor is parallel, $\nabla {\rm Ric}\equiv\mathring{\nabla} {\rm Ric}=0$, which is equivalent to the condition \eqref{eq:Ricparalaxial}. Then the equations of motion of the Lagrangian \eqref{eq:LagKV} are satisfied.
\end{theorem}

It is worth remarking that the Lagrangian we are considering to derive this result is just the Yang-Mills-like restriction \eqref{eq:LagKV}. However, during the elaboration of this review we noticed that the previous theorem can be improved as follows:
\begin{theorem}
For the geometry described in Theorem \ref{thm:Ax} (with parallel Ricci) the equation of the coframe is fulfilled for the entire strong sector \eqref{eq:LagStrong}, whereas the equation of the connection is proportional to $v_2 \mathcal{A}(\varphi) \mathcal{A}^\prime(\varphi)$.
\end{theorem}
In other words, the solution of \cite{PasicBarakovic2015} can be extended to arbitrary strong sectors with $v_2=0$. This is a new result, as far as the author of the present text is concerned.

\section{An Ansatz to search for general solutions}\label{sec:Obu}

Here we will proceed with the findings in \cite{ObukhovJCA2021}. This paper generalized the results in \cite{Obukhov2006} (also in MAG) and the solutions in Poincaré gravity in \cite{PGwave1}.

\subsection{Theory and Ansatz}

In this section we consider the full (even-parity) quadratic MAG action \eqref{eq:LAG0}. The metric will be again a Brinkmann metric with flat spatial slices, namely a pp-wave \eqref{eq:ppw}. For the connection we consider
\begin{equation}
  \dfGa_a{}^b = -\dfk(k_a V^b+k^b W_a)+k_ak^b U_a\cofr^a\,,\label{eq:ConObu}
\end{equation}
where $\dfk=k_a\cofr^a=\dex u$ is the wave 1-form of this geometry and the three vector variables $W^a$, $V^a$ and $U_a$ are assumed to be transversal and only dependent on $u$ and the transversal coordinates. Symbolically,
\begin{equation}
  U_a= \delta_a^A U_A(u,x,y),\qquad W^a= \delta_A^a W^A(u,x,y),\qquad V^a= \delta_A^a V^A(u,x,y)\,.
\end{equation}

\subsection{Solutions}

After plugging the Ansatz into the field equations, it can be proved that the resulting system of differential equations is of the form:
\begin{align}
[\text{EoM}~\cofr]^a\quad0 & =(...)\underline{\Delta}H+(...)\partial_{A}W^A+(...)\partial_{A}V^A+(...)\partial_A\uU{A}\,,\\
[\text{EoM}~\dfGa]_{[ab]}\quad0 & =(...)\partial_{A}H+(...)\uW{A}+(...)\uV{A}+(...)U_A\nonumber \\
 & \quad-\frac{\kappa}{4\rho}\Big\{\underline{\Delta}\big[(...)\uW{A}+(...)\uV{A}\big]+\partial_A\partial_B\big[(...) W^B +(...)V^B\big] \Big\}\nonumber \\
 & \quad-\frac{\kappa}{4\rho}\epsilon_{AB}\underline{\partial}^B\left\{ \epsilon^{CD}\partial_{[C}\big[(...) \uW{D]}+(...) \uV{D]}+ (...)U_{D]}\big]\right\} \,,\label{eq:eomObu}\\
[\text{EoM}~\dfGa]_{(ab)}\quad0 & =(\text{same structure as \eqref{eq:eomObu}}),\\
0 & =(\text{same structure as \eqref{eq:eomObu} without the second line}),\\
0 & =\partial_u \left[(...)\partial_{A} W^{A}+ (...)\partial_{A} V^A +(...)\partial_A \uU{A} \right]\,.
\end{align}
where (...) means {\it a certain combinations of parameters from the action} (for the exact value of these coefficients, check \cite{ObukhovJCA2021}). As we can see, three of the equations have a free transversal index and the other two have no free indices. Here we are using the Levi-Civita symbol $\epsilon_{AB}$ defined by $\epsilon_{23}:=1$ and the shortcuts
\begin{align}
  \uW{A} &:= \delta_{AB}W^B,\qquad  
  &\uV{A} &:= \delta_{AB}V^B,\qquad
  &\uU{A} &:= \delta^{AB}U_B,\nonumber\\
  \underline{\partial}^{A}  &:= \delta^{AB}\partial_B ,\qquad
  &\underline{\Delta} &:= \delta^{AB}\partial_A \partial_B.
\end{align}

In order to get closer to the solutions, the following procedure can be followed:
\begin{enumerate}
\item First, we perform a potential-copotential decomposition of the three unknown transversal vectors:
    \begin{align}
      W^A &=: \frac{1}{2}\Big( \delta^{AB}\partial_B\mathcal{W} + \epsilon^{AB}\partial_B\overline{\mathcal{W}} \Big),\\
      V^A &=: \frac{1}{2}\Big( \delta^{AB}\partial_B\mathcal{V} + \epsilon^{AB}\partial_B\overline{\mathcal{V}} \Big),\\
      U_A &=: \frac{1}{2}\Big( \partial_A\mathcal{U} + \epsilon_{AB}\delta^{BC}\partial_C\overline{\mathcal{U}}\Big).
    \end{align}
After this redefinition, the six degrees of freedom are encoded into the three potential variables $\{\mathcal{W}, \mathcal{V}, \mathcal{U}\}$ (even parity) and the three copotentials $\{\overline{\mathcal{W}}, \overline{\mathcal{V}}, \overline{\mathcal{U}}\}$ (odd parity). This allows to simplify considerably the dynamical equations, thanks to the properties
  \begin{equation}
    F^{A}=\frac{1}{2}\Big(\delta^{AB}\partial_{B}\mathcal{F}+\epsilon^{AB}\partial_{B}\overline{\mathcal{F}}\Big)\qquad\Rightarrow\qquad
    \left\{ \begin{array}{rl}
      2\partial_{A}F^{A} & =\underline{\Delta}\mathcal{F}\\
      2\epsilon^{AB}\partial_{[A}F_{B]} & =\underline{\Delta}\overline{\mathcal{F}}
    \end{array}\right.\,. 
  \end{equation}

\item The vector equations (those with a free transversal index) can also be decomposed in a similar way:
\begin{equation}
  0=\mathbf{E}^{A}\equiv\delta^{AB} \partial_{B}\mathbf{E}+\epsilon^{AB}\partial_{B}\overline{\mathbf{E}}\,.
\end{equation}
This is true if and only if $\mathbf{E}=\xi$ and $\overline{\mathbf{E}}= \overline{\xi}$, where  $\xi$ and $\overline{\xi}$ are arbitrary harmonic functions in the transversal flat directions ({\it i.e.}, such that $\underline{\Delta} \xi=0=\underline{\Delta}\overline{\xi}$). It can be shown that, thanks to the linearity of the equations, it is possible to redefine the potentials and copotentials in order to absorb all of these harmonic functions so that one can just impose
  \begin{equation}
    \mathbf{E}=0\qquad \text{and}\qquad  \overline{\mathbf{E}}=0.
   \end{equation}
This discussion was also done in \cite{PGwave13} for waves in Poincaré Gauge gravity. Due to the fact that the Lagrangian only contains even-parity invariants, the system of equations coming from $\mathbf{E}$'s exclusively depends on the even-parity variables $\{H,\, \mathcal{W} ,\,\mathcal{V},\,\mathcal{U} \}$, whereas the one coming from $\overline{\mathbf{E}}$'s only contains the copotentials $\{\overline{\mathcal{W}},\, \overline{\mathcal{V}},\, \overline{\mathcal{U}}\}$. Therefore, even and odd variables get decoupled into two separated systems that can be solved independently.

\item We choose a new (more appropriate) set of variables.
    \begin{equation}
      \begin{array}{rl}
        \mathcal{X}_{0} & =\mathcal{W}-\mathcal{V}\\
        \mathcal{X}_{1} & =H -\mathcal{W}+\mathcal{U}\\
        \mathcal{X}_{2} & =\mathcal{W}+\mathcal{V}+\mathcal{U}\\
        \mathcal{X}_{3} & =\mathcal{W}+\mathcal{V}-2\mathcal{U}
      \end{array}\quad\Leftrightarrow\quad\begin{array}{rl}
        H & =\frac{1}{2}\mathcal{X}_{0}+\mathcal{X}_{1}+\frac{1}{2}\mathcal{X}_{3}\\
        \mathcal{W} & =\frac{1}{2}\mathcal{X}_{0}+\frac{1}{3}\mathcal{X}_{2}+\frac{1}{6}\mathcal{X}_{3}\\
        \mathcal{V} & =-\frac{1}{2}\mathcal{X}_{0}+\frac{1}{3}\mathcal{X}_{2}+\frac{1}{6}\mathcal{X}_{3}\\
        \mathcal{U} & =\frac{1}{3}\mathcal{X}_{2}-\frac{1}{3}\mathcal{X}_{3}
      \end{array}
    \end{equation}
    \begin{equation}
      \begin{array}{rl}
        \overline{\mathcal{X}}_{1} & =-\overline{\mathcal{W}}+\overline{\mathcal{U}}\\
        \overline{\mathcal{X}}_{2} & =\overline{\mathcal{W}}+\overline{\mathcal{V}}+\overline{\mathcal{U}}\\
        \overline{\mathcal{X}}_{3} & =\overline{\mathcal{W}}+\overline{\mathcal{V}}-2\overline{\mathcal{U}}
      \end{array}\quad\Leftrightarrow\quad\begin{array}{rl}
        \overline{\mathcal{W}} & =-\,\overline{\mathcal{X}}_{1}+\frac{1}{3}\overline{\mathcal{X}}_{2}-\frac{1}{3}\overline{\mathcal{X}}_{3}\\
        \overline{\mathcal{V}} & =\overline{\mathcal{X}}_{1}+\frac{1}{3}\overline{\mathcal{X}}_{2}+\frac{2}{3}\overline{\mathcal{X}}_{3}\\
        \overline{\mathcal{U}} & =\frac{1}{3}\overline{\mathcal{X}}_{2}-\frac{1}{3}\overline{\mathcal{X}}_{3}.
      \end{array}
    \end{equation}

\item Finally, in \cite{ObukhovJCA2021} a plane-wave dependency of the type
    \begin{equation}
      \mathcal{X}_I=\mathcal{X}^{(0)}_I(u) {\rm e}^{{\rm i} q_A x^A}\,, \qquad \overline{\mathcal X}_I=\overline{\mathcal X}^{(0)}_I(u) {\rm e}^{{\rm i} \overline{q}_A x^A}\,,\label{eq:X0}
    \end{equation}
was assumed to reduce the differential operators in the transversal directions into algebraic operators. The resulting equations are the same as those obtained by performing a Fourier transform with respect to these transversal directions,
    \begin{equation}
    \mathcal{X}_I= \int \hat{\mathcal{X}}_I(u, q_B) {\rm e}^{{\rm i} q_A x^A} \dex^2q\,, \qquad \overline{\mathcal X}_I = \int \hat{\overline{\mathcal X}}_I(u, \overline{q}_B) {\rm e}^{{\rm i}\overline{q}_A x^A} \dex^2\overline{q},
\end{equation}
assuming that $\{\mathcal{X}_I\}$ and $\{\overline{\mathcal X}_I\}$ as well as their derivatives decay sufficiently fast at the transversal infinity. Therefore, although \eqref{eq:X0} seems to be a restriction, what we get from it are just the equations of motion in the transversal Fourier space.
\end{enumerate}

With this procedure, the equations of motion can be expressed as one restriction on the $u$-dependency of the even amplitudes
 \begin{equation}
      v_4\,\partial_u\hat{\mathcal X}{}_0 + {\frac 23}(\varpi_0-6z_1)\,\partial_u \hat{\mathcal X}{}_2 + {\frac 13}\varpi_0\,\partial_u\hat{\mathcal X}{}_3 = 0,\label{eq:restu}
    \end{equation}
and seven additional equations that can be rearranged as
\begin{equation}
  \mathrm{M} \left(\begin{array}{c}\hat{\mathcal X}_0 \\ \hat{\mathcal X}_1 \\
      \hat{\mathcal X}_2 \\ \hat{\mathcal X}_3\end{array}\right) =  0,\qquad \overline{\mathrm{M}}\left(\begin{array}{c}\hat{\overline{\mathcal X}}{}_1 \\
      \hat{\overline{\mathcal X}}{}_2 \\ \hat{\overline{\mathcal X}}{}_3\end{array}\right) =  0,\label{eq:MMbareqs}
\end{equation}
with
\renewcommand\arraystretch{1.5}
\begin{align}
      \mathrm{M}&\coloneqq\left(\begin{array}{cccc} -{\frac {a_0}{2}} & a_1 & 0 & c_1 \\
      0 & 2c_1 - a_1 & {\frac 13}(a_0 - 4b_1) & {\frac {a_0}{6}} - c_1 + {\frac 43}b_2 \\
      2(w_1 + w_4){\mathcal Q}^2 & a_0 + a_1 & {\frac 23}v_4{\mathcal Q}^2 &
      {\frac {a_0}{2}} + c_1 + {\frac 13}v_4{\mathcal Q}^2 \\
      v_4{\mathcal Q}^2 & 0 & a_0 - 4b_1 + {\frac 23}\varpi_0{\mathcal Q}^2 &
      {\frac 13}{\mathcal Q}^2\varpi_0 \end{array}\right),
     \label{eq:Matrixeven}\\
     \overline{\mathrm{M}}&\coloneqq \left(\begin{array}{ccc} 0 & a_0 - 4b_1 + 4z_1\overline{\mathcal Q}{}^2 & 0 \\
      a_0 + 2c_1 - 2\varpi_2\overline{\mathcal Q}^2 & 0 & \frac{2(a_0 + 2b_2)}{3}
      - (\varpi_2 + \varpi_3)\overline{\mathcal Q}{}^2\\
      a_0 + a_1 -\varpi_1\overline{\mathcal Q}{}^2 & 0 & {\frac {a_0}{2}} + c_1
      - \varpi_2\overline{\mathcal Q}{}^2\end{array}\right). \label{eq:Matrixodd}
\end{align}
\renewcommand\arraystretch{1}
In these expressions we are using the abbreviations:
  \begin{align}
      \varpi_0 &:= 3z_1 + z_4 + 2v_4\,,\ &\varpi_1 &:= 4(w_1 + w_2),\nonumber\\ 
      \varpi_2 &:= 2(w_1 + w_2) + v_2,\  &\varpi_3 &:= {\frac 13}(z_1 + 3z_2)\,
    \end{align}
and 
\begin{equation}
  {\mathcal Q}^2  :=  {\frac{\kappa}{4\rho} q_Aq_B\delta^{AB}}\,,\qquad\overline{\mathcal Q}{}^2 := {\frac{\kappa}{4\rho}\overline{q}_A\overline{q}_B\delta^{AB}}\,.
\end{equation}
We can now state the following:
\begin{theorem}
Consider the configuration given by a pp-wave metric \eqref{eq:ppw} and the connection \eqref{eq:ConObu}. The solutions of this type for the general action \eqref{eq:LAG0} are determined by the solutions of the system of equations \eqref{eq:restu}-\eqref{eq:MMbareqs}.
\end{theorem}

From this system of equations one can try to find solutions for different sets of parameters. In order to get generic non-trivial solutions we just need to ensure that the determinants of the matrices in \eqref{eq:Matrixeven} and \eqref{eq:Matrixodd} are zero. In \cite{ObukhovJCA2021}, particular solutions of the Riemannian, teleparallel and pseudo-instanton type were provided. The main interest of this derivation is that the resulting equations are valid for arbitrary parameters in the quadratic action.

\section{Miscellanea: Riemannian solutions and colliding waves}\label{sec:coll}

Regarding Riemannian solutions, in \cite{ObukhovJCA2021} it was proven that for generic parameters of the full quadratic action \eqref{eq:LAG0}, the solution of General Relativity is a solution of the theory (in vacuum). In terms of the objects introduced in Section \ref{sec:Obu}, the Riemannian constraint is
\begin{equation}
  W^A=-V^A=\frac{1}{2}\delta^{AB}\partial_B H. \label{eq:GRsol1}
\end{equation}
Moreover, the solution of General Relativity additionally verifies the harmonic condition \eqref{eq:derriccihar}, namely,
\begin{equation}
  \underline{\Delta} H=0.\label{eq:GRsol2}
\end{equation}
For the specific set of parameters $v_4=w_1+w_4=a_0=0$, a solution is obtained for arbitrary $H$. Interestingly, as was mentioned in \cite{ObukhovJCA2021} and under their particular Ansatz, the General Relativity solution \eqref{eq:GRsol1}-\eqref{eq:GRsol2} is not just one solution but the general one for the weak gravity theory with arbitrary parameters \eqref{eq:LagWeak}, except for those that give rise to the Einstein-Hilbert term,\footnote{Here, as in \cite{JCAThesis}, we correct misprints in the original publication \cite{ObukhovJCA2021}.}
\begin{equation}
a_0= -\,a_1 = {\frac {a_2}{2}} = 2a_3 = 4b_1 = -2b_2 = - 8b_3 = {\frac {8b_4}{3}} = 2b_5 = -2c_1 = c_2 = c_3 \,.
\end{equation}
In such a special case, not only the Levi-Civita connection (given by \eqref{eq:GRsol1}) but any other connection together with \eqref{eq:GRsol2} gives a solution of the theory.\footnote{
    The reason is that the Lagrangian reduces to just the Einstein-Hilbert term, but the theory is metric-affine by construction. Hence, the equation of motion of the connection gives an identity ($0=0$) which is verified by all connections.}

In addition to these results, it is also worth remarking that under very generic considerations all the Riemannian solutions of the strong sector \eqref{eq:LagStrong} can be separated into three types: Einstein spaces, local products of two Einstein 2-manifolds and pp-waves \eqref{eq:ppw}, as was shown in \cite{Vassiliev2005}. Here by ``very generic considerations'' we mean that, to prove this result, the parameters of the strong sector have to respect some inequalities (see Theorem 7.1 in \cite{Vassiliev2005}).

~

\begin{figure}
\begin{center}
\includegraphics[width=\textwidth]{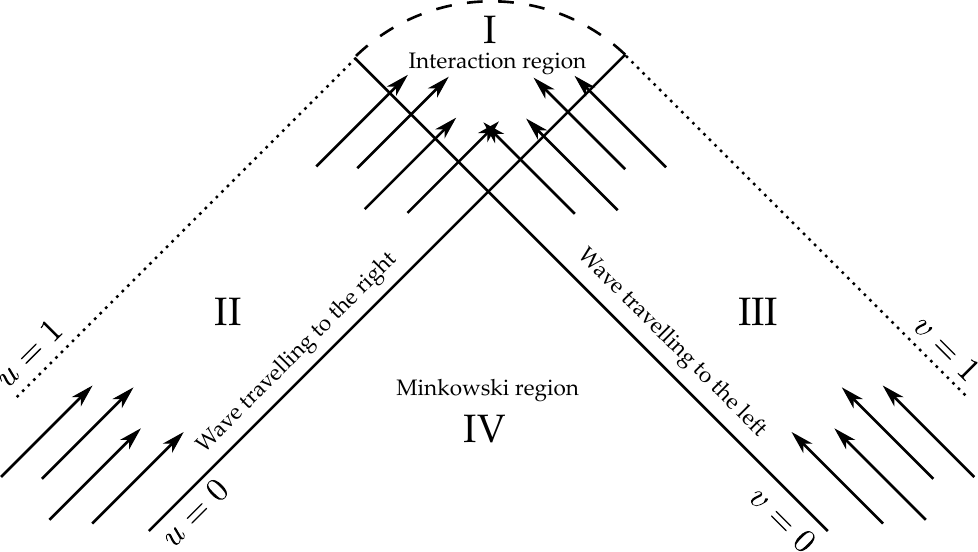}
\end{center}
\caption{Diagram of two colliding pp-waves. Region IV has a flat metric, whereas regions II and III correspond to pp-wave geometries that approach each other with $u=0$ and $v=0$ as wavefronts, respectively. In the point $(u,v)=(0,0)$ of the diagram, the collision takes place and I is the interaction region after it.
}\label{fig:colgw}
\end{figure}

Finally, we also comment on the colliding wave\footnote{
    See \cite{GriffCol} for a more detailed review on colliding wave geometries in General Relativity.}
configurations found in \cite{Garcia1998, MaciasLammerzahl2000}. The authors made use of the Obukhov's equivalence theorem to get the gravitational part of the equations of motion in the Einstein-Proca form with $m=0$, as described in Section \ref{sec:Puetz}. The considered metric in the interaction region has the form
\begin{equation}
  \dex s^2 = g_{uu}(u,v)\dex u^2 + g_{AB}(u,v) \dex x^A \dex x^B\,,
\end{equation}
and for the triplet 1-form (defined in \eqref{eq:triplet}) in the same region it is assumed
\begin{equation}
  \dfphi=\dfphi(u,v)\,.
\end{equation}
Then the solution can be extended to the rest of the manifold consistently by performing a distributional substitution $(u,v)\to(u\Theta(u),v\Theta(v))$, where $\Theta$ is the Heaviside step function $\Theta(x)=1$ for $x>0$ and 0 elsewhere. For the explicit expressions, see the original publications \cite{Garcia1998, MaciasLammerzahl2000}. In \cite{Garcia1998}, a vacuum solution was found, and a Maxwell field was incorporated in \cite{MaciasLammerzahl2000}. In these works, the regions $\{0<u<1,v<0\}$ and $\{u<0,0<v<1\}$  (regions II and III in Figure \ref{fig:colgw}) describe type-N wave propagation with a non-vanishing covariantly constant null vector, implying that the metric admits a Brinkmann form \eqref{eq:Brink} after an appropriate coordinate transformation. The interaction region is governed by a type-D axially symmetric geometry.\footnote{
    In the context of Einstein-Maxwell theory, there is a one-to-one correspondence between solutions with axial symmetry and colliding wave solutions \cite{Steph}.}
Although the properties of the singular boundary were not discussed in the original publications, due to the mutual focusing of the waves a curvature singularity usually appears in the future of the interaction region $\{u>0,v>0\}$ (dashed line in Figure \ref{fig:colgw}) as, for instance, in the Khan-Penrose geometry \cite{Griff, GriffCol}. The points in $\{u=1,v<0\}$ and $\{u<0,v=1\}$ (dotted lines in Figure \ref{fig:colgw}) then become the so-called fold singularities \cite{Griff, GriffCol}.

\section{Overview and final comments}\label{sec:fin}

The most relevant results on non-Riemannian wave solutions have been collected in Theorems (from 3 to 9) throughout the manuscript. Now we proceed with a general overview of results.

Our starting point was a revision of fundamental notions on metric-affine geometry, including its basic objects and the derivation of the equations of motion for a general MAG theory with matter. Then we focused on the even-parity quadratic MAG Lagrangian and gave the explicit expressions for the gravitational momenta, which determine the gravitational part of the equations. Moreover, the particular case of the purely strong sector is also discussed. We then reviewed the gravitational wave metrics that are relevant for our analysis; in particular, we concentrated on Kundt spaces of type-N and Brinkmann solutions of the pp-wave type. We also provided information about the irreducible components of the curvature.

After this, we started the analysis of solutions. The first ones we covered were those in the presence of matter. In particular, we began with pp-wave geometries coupled to a Dirac spinor field and checked how the axial spinor current acts as a source of torsion \cite{TuckerWang1995}. Then, we studied the electrovacuum solutions in \cite{Garcia2000, Puetzfeld2002, PuetzfeldDip}, which belong to the more general class of Kund spaces described by \eqref{eq:genmet}; in this analysis, the triplet Ansatz together with the vanishing of a certain combination of parameters turn the equation of the connection into a Maxwell equation to be fulfilled by the triplet 1-form. The contribution of the latter to the equation of the coframe is similar to the one of the electromagnetic 1-form: they act as inhomogeneous terms in the differential equation of $\mathcal{H}$ (see \eqref{eq:EQHGar}).

At this point we turned to the vacuum solutions in \cite{KingVassiliev2001,Vassiliev2002,Vassiliev2005}, whose metric is assumed to be Minkowski and the torsion is the only non-trivial geometric structure. Then we covered the solutions in \cite{PasicVassiliev2005, PasicPhDThesis, PasicBarakovic2014, PasicBarakovic2015}, which involve pp-wave metric configurations and  metric-compatible connection. Solutions with non-trivial purely tensoral torsion were found for the general strong sector after imposing the condition of parallel Ricci tensor \cite{PasicVassiliev2005, PasicPhDThesis, PasicBarakovic2014}. A similar analysis but with an axial torsion was done in \cite{PasicBarakovic2015}. Their results only applied to the restricted Lagrangian \eqref{eq:LagKV}, but, as we mentioned in this review, they can be extended to arbitrary strong sectors with $v_2=0$.

We continued with the solutions with both torsion and nonmetricity discovered in \cite{ObukhovJCA2021}, which generalize those of \cite{Obukhov2006}. We explained the procedure to simplify the equations, the splitting of even and odd variables and the transformation of the transversal derivatives in the equations into algebraic terms. This allows a simpler analysis of general solutions as discussed in the original publication.

Finally, we reviewed some results concerning Riemannian geometries of the pp-wave type in metric-affine gravity. In addition to this, we also included some comments about the colliding wave solutions in \cite{Garcia1998, MaciasLammerzahl2000} and their structure. 

To finish this review we remark that there are many open areas that need further research regarding gravitational wave exact solutions. A good example could be solutions with in-falling matter, which might describe interesting physical scenarios involving the matter microstructure. The addition of odd-parity contributions in the action is also a relevant field of study that gives a more complicated dynamics, coupling the even-parity and the odd-parity parts of the fields; as a result, different dispersion relations are expected for different polarizations in the metric and/or the connection propagating functions. Another important aspect to study is the presence of a non-trivial cosmological constant term in the Lagrangian, a setting that requires more complicated Ansatzes for the coframe.

\section*{Acknowledgments}

The author would like to thank Yuri Obukhov, Dirk Puetzfeld and Jorge Gigante Valcarcel for their useful comments and feedback. This research was supported by the European Regional Development Fund through the Center of Excellence TK133 ``The Dark Side of the Universe'' and by the Mobilitas Pluss postdoctoral grant MOBJD1035.

\appendix

\section{Explicit expressions for the irreducible components}\label{app:irred}

In this review we follow the conventions and notation in \cite{ObukhovJCA2021}, which we reproduce here for completeness in arbitrary dimension $\dimM$. For the expressions in components, see Appendix B of \cite{JCAThesis}.

\subsection{Decomposition of the torsion and the nonmetricity}

The three irreducible parts of the torsion described after \eqref{eq:irredT} are given by
\begin{align}
  \irrdfT{2}{}^a & \coloneqq\frac{1}{\dimM-1}\cofr^a\wedge\dfT\,,\\
  \irrdfT{3}{}^a & \coloneqq\frac{1}{3}\dint{\vfre^a}(\dfT^b\wedge\cofr_b)\,,\\
  \irrdfT{1}{}^a & \coloneqq\dfT^a-\irrdfT{2}{}^a -\irrdfT{3}{}^a.
\end{align}
For the nonmetricity we have the expressions:
\begin{align}
  \irrdfQ{4}{}_{ab} &\coloneqq g_{ab}\dfQ\,,\\
  \irrdfQ{3}{}_{ab} &\coloneqq \frac{2\dimM}{(\dimM-1)(\dimM+2)}\left(\cofr_{(a}\dint{\vfre_{b)}}\dfLa-\frac{1}{\dimM}g_{ab}\dfLa\right)\,,\\
  \irrdfQ{2}{}_{ab} & \coloneqq-\frac{2}{3}{\rm sgn}(g)\star\left(\overline{\dfLa}_{(a}\wedge\cofr_{b)}\right)\,,\\
  \irrdfQ{1}{}_{ab} & \coloneqq\dfQ_{ab}-\irrdfQ{2}{}_{ab}-\irrdfQ{3}{}_{ab}-\irrdfQ{4}{}_{ab}\,,
\end{align}
where we have introduced the auxiliary $(\dimM-2)$-form
\begin{equation}
  \overline{\dfLa}_a \coloneqq\star\left[(\dfQ_{ac}-g_{ac}\dfQ) \land\cofr^{c}-\frac{1}{\dimM-1}\cofr_a\wedge\dfLa\right]\,.
\end{equation}
The symbol ${\rm sgn}(g)$ is the sign of the determinant of the metric. For Lorentzian metrics in four dimensions, ${\rm sgn}(g)=-1$.

\subsection{Decomposition of the curvature}

For the curvature 2-form we first split it into symmetric and antisymmetric parts (in the external indices):
\begin{align}
  \dfR_{ab} & =\dfW_{ab}+\dfZ_{ab}\,,
\end{align}
where $\dfW_{ab}\coloneqq\dfR_{[ab]}$ and $\dfZ_{ab}\coloneqq\dfR_{(ab)}$.

Under the pseudo-orthogonal group, the antisymmetric part can be separated into six irreducible parts $\dfW_{ab}=\sum_{I=1}^{6}\irrdfW{I}{}_{ab}$, given by
\begin{align}
  \irrdfW{2}{}_{ab} & \coloneqq{\rm sgn}(g)\star\left(\cofr_{[a}\wedge\overline{\dfPsi}{}_{b]}\right),\\
  \irrdfW{3}{}_{ab} & \coloneqq{\rm sgn}(g)\frac{1}{12}\star\left(\overline{\dfX}\wedge\cofr_{ab}\right),\\
  \irrdfW{4}{}_{ab} & \coloneqq-\frac{2}{\dimM-2}\cofr_{[a}\wedge\dfPsi_{b]},\\
  \irrdfW{5}{}_{ab} & \coloneqq-\frac{1}{\dimM-2}\cofr_{[a}\wedge\dint{\vfre_{b]}}\left(\cofr^c\wedge\dfX_c\right),\\
  \irrdfW{6}{}_{ab} & \coloneqq-\frac{1}{\dimM(\dimM-1)}X\cofr_{ab},\\
  \irrdfW{1}{}_{ab} & \coloneqq\dfW_{ab}-\sum_{\mathrm{I}=2}^{6}\irrdfW{I}{}_{ab},
\end{align}
where
\begin{align}
 \dfX^a              & \coloneqq\dint{\vfre_b}\dfW^{ab}\,, &  
 \overline{\dfX}{}^a   & \coloneqq\star(\dfW^{ba}\wedge\cofr_b)\,,\nonumber \\
 X                   & \coloneqq\dint{\vfre_a}\dfX^a, &  
 \overline{\dfX}       & \coloneqq\dint{\vfre_a}\overline{\dfX}{}^a\,,
\end{align}
\begin{align}
   \dfPsi{}_a          & \coloneqq\dfX_a -\frac{1}{\dimM}X\cofr_a -\frac{1}{2}\dint{\vfre_a}(\cofr^b \land\dfX_b)\,,\nonumber\\
   \overline{\dfPsi}{}_a & \coloneqq\overline{\dfX}{}_a-\frac{1}{4}\cofr_a \land\overline{\dfX} -\frac{1}{\dimM-2}\dint{\vfre_a}(\cofr^b \land\overline{\dfX}{}_b) \,.
\end{align}

Moreover, the symmetric part contains five irreducible parts $\dfZ_{ab}=\sum_{I=1}^{5}\irrdfZ{I}{}_{ab}$ defined as follows
\begin{align}
  \irrdfZ{2}{}_{ab} & \coloneqq\frac{1}{2}{\rm sgn}(g)\star\left(\cofr_{(a}\wedge\overline{\dfPhi}{}_{b)}\right),\\
  \irrdfZ{3}{}_{ab} & \coloneqq\frac{1}{\dimM^{2}-4}\left[\dimM\cofr_{(a}\wedge\dint{\vfre_{b)}}(\cofr^c\wedge\dfY_c)-2g_{ab}(\cofr^c\wedge\dfY_c)\right],\\
  \irrdfZ{4}{}_{ab} & \coloneqq\frac{2}{\dimM}\cofr_{(a}\wedge\dfPhi{}_{b)},\\
  \irrdfZ{5}{}_{ab} & \coloneqq\frac{1}{\dimM}g_{ab}\dfZ_c{}^c\qquad =\frac{1}{2}g_{ab}\dex \dfQ \label{eq:Z5},\\
  \irrdfZ{1}{}_{ab} & \coloneqq\dfZ_{ab}-\sum_{\mathrm{I}=2}^{5}\irrdfZ{I}{}_{ab}, 
\end{align}
where we have introduced
\begin{align}
\dfY_a             & \coloneqq\dint{\vfre^b}(\dfZ_{ab}-\irrdfZ{5}{}_{ab})\,, &  
\overline{\dfY}{}_a  & \coloneqq\star[(\dfZ_{ba}-\irrdfZ{5}{}_{ba})\wedge\cofr^b]\,,\nonumber \\
\dfPhi{}_a         & \coloneqq\dfY_a-\frac{1}{2}\dint{\vfre_a}(\cofr^b\wedge\dfY_b)\,, &  
\overline{\dfPhi}{}_a& \coloneqq\overline{\dfY}{}_a-\frac{1}{\dimM-2}\dint{\vfre_a}(\cofr^b\wedge\overline{\dfY}{}_b)\,.
\end{align}

Indeed, the object $\dfP$ introduced in \eqref{eq:PPXYZ} can be expressed in terms of the previously defined objects as
\begin{equation}
  \dfP = -\cofr^a\wedge \dfX_a + \cofr^a\wedge \dfY_a + {\frac 12}\dfZ_a{}^a\,.
\end{equation}

Finally we collect some remarks about particular cases:
\begin{itemize}
  \item For a metric-compatible connection all the $\irrdfZ{I}_{ab}$ are zero.

  \item For the Levi-Civita connection only the Weyl part, the traceless-symmetric Ricci and the Ricci scalar parts survive, respectively, $\irrdfWLC{1}_{ab}$, $\irrdfWLC{4}_{ab}$ and $\irrdfWLC{6}_{ab}$.

  \item For a connection exclusively given by the traces $\dfT$, $\dfQ$ and $\dfLa$ (the rest of the irreducible parts of the torsion and the nonmetricity are assumed to be zero), we have
\begin{equation}
\irrdfW{2}_{ab}=\irrdfW{3}_{ab}=\irrdfZ{1}_{ab}=\irrdfZ{2}_{ab}=0\,,\qquad \irrdfW{1}_{ab}=\irrdfWLC{1}_{ab}\,.
\end{equation}
The remaining ones, $\irrdfW{4}_{ab}$, $\irrdfW{5}_{ab}$ $\irrdfW{6}_{ab}$, $\irrdfZ{3}_{ab}$ and $\irrdfZ{4}_{ab}$ generically depend on the three 1-forms, $\dfT$, $\dfQ$ and $\dfLa$. However, the coefficients of all the $\dfQ$-dependent terms in the expression of $\irrdfZ{3}_{ab}$ are proportional to $(\dimM-4)$; hence, if we focus on the four-dimensional case, $\irrdfZ{3}_{ab}$ becomes exclusively $(\dex\dfLa)$-dependent, playing the role of a sort of field strength for the 1-form $\dfLa$. The triplet Ansatz \eqref{eq:triplet} is a particular case of this geometry.
\end{itemize}

\end{document}